\documentclass[preprint,5p,twocolumn]{elsarticle}

\usepackage{amssymb}
\usepackage{amsmath}
\usepackage{xcolor}
\usepackage{enumitem}
\usepackage{fancyvrb}
\usepackage{threeparttable}
\usepackage{hyperref}
\newcounter{bla}

\journal{Computer Physics Communications}

\begin{document}

\begin{frontmatter}



\title{pCI: a parallel configuration interaction software package for high-precision atomic structure calculations}


\author[a]{Charles Cheung\corref{cor1}}
\ead{ccheung@udel.edu}
\author[b,c]{Mikhail G. Kozlov}
\author[a]{Sergey G. Porsev}
\author[a]{Marianna S. Safronova}
\author[d]{Ilya I. Tupitsyn}
\author[e,f]{Andrey I. Bondarev}

\cortext[cor1] {Corresponding author}
\address[a]{Department of Physics and Astronomy, University of Delaware, Delaware 19716, USA}
\address[b]{Petersburg Nuclear Physics Institute of NRC ``Kurchatov Institute'', Gatchina, Leningrad District 188300, Russia}
\address[c]{St. Petersburg Electrotechnical University “LETI”, Prof. Popov Str. 5, St. Petersburg, 197376, Russia}
\address[d]{Department of Physics, St. Petersburg State University, Ulianovskaya 1, Petrodvorets, 198504 St. Petersburg, Russia}
\address[e]{Helmholtz Institute Jena, 07743 Jena, Germany}
\address[f]{GSI Helmholtzzentrum f\"ur Schwerionenforschung GmbH, 64291 Darmstadt, Germany}

\begin{abstract}

We introduce the pCI software package for high-precision atomic structure calculations. The standard method of calculation is based on the configuration interaction (CI) method to describe valence correlations, but can be extended to attain better accuracy by including core correlations via many-body perturbation theory (CI+MBPT) or the all-order (CI+all-order) method, as well as QED corrections via QEDMOD. The software package enables calculations of atomic properties, including energy levels, $g$-factors, hyperfine structure constants, multipole transition matrix elements, polarizabilities, and isotope shifts. It also features modern high-performance computing paradigms, including dynamic memory allocations and large-scale parallelization via the message-passing interface, to optimize and accelerate computations.

\end{abstract}


\end{frontmatter}

% \linenumbers

% All CPiP articles must contain the following
% PROGRAM SUMMARY.

{\bf PROGRAM SUMMARY}

\begin{small}
\noindent
{\em Program Title:}
    pCI \\
{\em CPC Library link to program files:} (to be added by Technical Editor) \\
{\em Developer's repository link:} 
    \url{https://github.com/ud-pci/pCI} \\
{\em Code Ocean capsule:} (to be added by Technical Editor)\\
{\em Licensing provisions:} 
    GPLv3  \\
{\em Programming language:} 
    Fortran                                 \\
{\em Supplementary material:}
    Documentation available at https://pci.readthedocs.io \\
{\em Nature of problem:} 
    Calculation of atomic and ionic properties, including energy levels, hyperfine structure constants, and multipole transition matrix elements. \\
{\em Solution method:}
    The software package calculates energies and associated wave functions for the desired atomic states using the configuration interaction method. Using wave functions, different atomic properties can be obtained, including $g$-factors, hyperfine structure constants, transition amplitudes, and polarizabilities.  \\
{\em Unusual features:}
    One-electron orbitals outside the nucleus are defined on the radial grid points. Inside the nucleus, they are described in a Taylor expansion over $r/R$, where $R$ is the nuclear radius.  \\
{\em Restrictions:}
    This software package is not designed for calculations of high Rydberg states and continuous spectrum.  \\
{\em Additional comments:}
    All serial programs have been compiled and tested with the freely available Intel Fortran compilers ``ifort'', and all parallel programs with the OpenMPI wrapper ``mpifort'' for Intel Fortran compilers.

\end{small}

\section{Introduction}\label{sec:intro}
Accurate atomic theory is indispensable to the design and interpretation of a wide range of experiments, with direct experimental measurement of relevant parameters being infeasible or impossible. Many applications, ranging from studies of fundamental interactions to the development of future technologies, require precise knowledge of various atomic properties, such as energy levels, wavelengths, transition rates, branching ratios, lifetimes, hyperfine constants, polarizabilities, and others. The need for high-precision atomic modeling has increased significantly in recent years with the development of atom-based quantum technologies for a wide range of fundamental and practical applications.

Further rapid advances in applications involving complex atoms require accurate knowledge of basic atomic properties, most of which remain highly uncertain and difficult to measure experimentally. Moreover, the lack of reliable data hinders the search for further applications of rich and complex atomic structures. In summary, there is a demonstrated need for high-quality atomic data and software in several scientific communities.

We have developed and tested state-of-the-art relativistic atomic codes~\cite{Safronova08,sym13040621,KozPorSaf15,Cheung2020,CIall} capable of computing a wide variety of atomic properties for a large number of atoms and ions~\cite{Sr2023,Ac,UU+,Cr,2016Pb,2015Sr,2020Sr}, including negative ions~\cite{La-,Bi-} and highly charged ions~\cite{Cheung2020,2022Fe16+,2024Fe16+,2024Ni18+,Cf1517,Th35+,2018HCIRMP}.
The data computed by these codes are needed for a wide science community working in the fields of quantum information and simulation~\cite{Sr2023,2020Sr,SUN}, degenerate quantum gases~\cite{PhysRevA.89.012711,PhysRevLett.109.230802}, atomic clocks~\cite{2015Sr,PhysRevLett.133.023401,Ti}, precision measurements~\cite{Ac,Th35+,Cr,2016Pb}, studies of fundamental symmetries~\cite{2019YbLLI,PhysRevLett.120.133205,2015Natur.517..592P}, dark matter searches~\cite{banerjee2023oscillatingnuclearchargeradii,2022EPJQT...9...12B},  and many others.
Atomic data are also highly demanded by the astrophysics~\cite{2022Fe16+,2024Fe16+,2024Ni18+,kilonova}, plasma~\cite{PhysRevA.96.062506,2017NIMPB.408..118N}, and nuclear physics~\cite{hyp,PhysRevLett.121.213001,PhysRevLett.120.232503} communities. 
This work makes these codes user-friendly and available to the scientific community, with specific examples demonstrating the capabilities of the code package. 

Several \textit{ab initio} atomic structure codes have been developed for public use in the last few decades. The NIST MCHF/MCDHF database contains not only collections of transition data, but also several atomic structure packages, based on different relativistic theories and computational methods~\cite{NISTdb}. These include ATSP2K, a multiconfiguration Hartree-Fock (MCHF) + Breit Pauli atomic structure package~\cite{MCHF}, and GRASP2K, which implements the fully relativistic multiconfiguration Dirac-Hartree-Fock (MCDHF) method for large-scale calculations~\cite{grasp2K}. Other commonly used and documented atomic structure codes based on configuration interaction (CI) include CIV3, which calculates the CI wave functions and the electric-dipole oscillator strengths~\cite{CIV3}. SUPERSTRUCTURE calculates the bound-state energies and associated radiative data in $LS$-coupling and intermediate coupling~\cite{SUPERSTRUCTURE}. The COWAN code utilizes a semi-empirical approach based on the CI method, where orbitals can be rescaled using least-squares fits to experimental data~\cite{COWAN}. HULLAC uses CI and the parametric potential method to calculate atomic structure and cross sections for collision and radiative processes~\cite{HULLAC}. The ATOM computer program system describes the atomic structure and processes based on HF and the random phase-exchange approximation~\cite{ATOM,atoms10020052}. FAC calculates various atomic radiative and collision processes based on the relativistic CI method~\cite{FAC}. These codes were developed decades ago; some of them have some recent updates, including dynamic memory allocation and limited parallelization. 

Modern applications require a much higher accuracy for a wider range of atomic properties than can be calculated with older codes. The first steps have been taken recently with the release of the GRASP2018 and AMBiT software packages. The GRASP2018 package introduces parallelism using the Message Passing Interface (MPI) and other optimizations to the original GRASP2K programs~\cite{grasp2018}, and the AMBiT software package features hybrid MPI+OpenMP parallelism to take full advantage of modern HPC architectures~\cite{ambit}. The pCI software package released here is an adaptation of the CI-MBPT code package, which implements CI with many-body perturbation theory (MBPT). In addition to the pure CI and CI+MBPT methods, pCI extends its capabilities with the CI+all-order approach. We ported the previous serial code package for use on the latest HPC architectures, enabling efficient code execution on large-scale computation facilities, in order to treat a wider range of problems beyond the original code package's capabilities. 

The pCI software package has been extensively used for calculations of the properties of atomic systems with up to 60 valence electrons~\cite{Cheung2020,sym13040621}. The modern high-performance computing methodologies employed here significantly expand the class of problems that the original CI-MBPT~\cite{KozPorSaf15} and the CI-all-order package~\cite{CIall} were capable of solving. Since the conception of parallel programs, the pCI has been utilized to calculate the atomic properties of a wide range of atomic systems, including those of neutral atoms~\cite{Ac,Cr,Sr2023,UU+}, highly charged ions~\cite{Cheung2020,Cf1517,Th35+,2024Fe16+}, and negative ions~\cite{La-,Bi-}. The breadth of applications related to these computations spanned from astrophysics to the development of novel atomic clocks and tests of fundamental physics.  

The newly developed programs were designed to be backward compatible with the auxiliary codes of the CI-MBPT code package~\cite{KozPorSaf15}, that is, they work naturally with the programs that extend the CI method to CI+MBPT, as well as the auxiliary programs that allow manipulation of the CI space. Users who wish to utilize the auxiliary codes of the 2015 CI-MBPT code package can refer to Ref.~\cite{KozPorSaf15} for instructions. 

The pCI programs are also compatible with programs that extend the CI method with the linearized coupled-cluster (all-order) approach. The all-order package is a standalone set of programs that employs the linearized coupled-cluster method with single- and double excitations for atomic structure calculations. It is used to construct an effective Hamiltonian~\cite{CIall}, which the CI then uses instead of the bare Hamiltonian. 
The inclusion of the all-order package enables high-precision computation for atoms and ions with a few valence electrons, generally significantly improving accuracy compared to the CI-MBPT approach~\cite{CIall,Sr2023,2020Sr,SUN,PhysRevA.89.012711,PhysRevLett.109.230802,2015Sr,PhysRevLett.133.023401,Ti,Ac,Cr,2016Pb}.
The CI+all-order package was tested up to six valence electrons~\cite{Cr}. 
A Fortran 77 version of this package is included in the pCI distribution, while a modern Fortran 90 version will be released in a future work. The all-order part of the package also includes a new faster MBPT code which can be used as a standalone code to implement the CI+MPBT method, replacing the codes released with the 2015 CI-MBPT package. We recommend the user to use the new MBPT variant. 
pCI can also utilize some capabilities of the QEDMOD package of~\cite{QEDMOD} to include contributions from quantum electrodynamics (QED). A set of compatible QED-related programs is included in the pCI distribution.

The names of most programs, input files, and output files are identical with those of the 2015 CI-MBPT package. The only exception are the programs that now utilize MPI parallelism, which have been prepended with a ``p'' to distinguish between the serial and parallel programs. The input and output files of the core pCI programs will be described in Section~\ref{sec:programs}. 
The major changes to the serial programs have been modernization efforts from F77 to F90. 
pCI has been utilized with up to 2048 computing cores and up to 32 TB of memory on the developers' in-house computer clusters Caviness and DARWIN at the University of Delaware. It has achieved near-perfect linear scalability and efficiency with the number of processors~\cite{sym13040621}, and will be discussed in detail in Sec.~\ref{sec:scalability}.

The pCI software package is open source and is actively being developed and maintained on GitHub in the \texttt{dev} branch, with tested updates pushed to the \texttt{main} branch. It is recommended to pull the pCI distribution from the \texttt{main} branch. 
Users are welcome to contribute to the project by submitting issues or pull requests on GitHub. 

This paper is summarized as follows. 
First, the theory of the atomic structure and methods of computation is discussed. 
Next, an overview of the pCI software package, including instructions on how to install and compile the programs, is given. 
Detailed explanations of the core programs, including their purpose, input files, and output files, are given in the following section. 
Then, the scalability and efficiency of the parallel programs are showcased. 
An example of a standard pCI workflow is demonstrated by computing the energies and oscillator strengths of highly charged Fe$^{16+}$. This includes constructing the basis set, constructing configuration lists, computing energies, and then finally computing reduced electric-dipole ($E1$) transition matrix elements, which are used to calculate the 3C/3D oscillator strength ratio. 
Following this, pCI-py, a set of helper scripts written in Python, is introduced to automate the pCI workflow. This simplifies and improves the user experience by automating tedious tasks, such as writing input files for each program.
Finally, an example of using these pCI-py scripts is given to calculate energies, reduced $E1$ matrix elements, as well as static and dynamic polarizabilities of neutral Sr.

\section{Theory}\label{sec:theory}
A standard approach to many-electron systems is to divide all electrons into core and valence electrons. In this way, we can separate the electron-electron correlation problem into two parts: one describing the valence-valence correlations under the frozen-core approximation, and the other describing the core-core and core-valence correlations. 

In the initial approximation, we start from the solution of the restricted Hartree-Fock-Dirac (HFD) equations in the central-field approximation to construct one-electron orbitals for the core and valence electrons. Virtual orbitals can be constructed from B-splines or other means to account for correlations. The valence-valence correlation problem is solved using the CI method, while core-core and core-valence correlations are included using either MBPT or the all-order method. In either case, an effective Hamiltonian is formed in the CI valence space, then diagonalized as in the usual CI method to find energies and wave functions for the low-lying states~\cite{KozPorSaf15}. 

In the valence space, the CI wave function is constructed as a linear combination of all distinct states of a specified angular momentum $J$ and parity.

\begin{equation}
    \Psi_J=\sum_i c_i \Phi_i, 
\end{equation}
where the set $\left\{\Phi_i\right\}$ are Slater determinants, enumerated by the index $i$, generated by exciting electrons from a set of reference configurations to higher orbitals. Varying the coefficients $c_i$ results in an eigenvalue problem

\begin{equation}
    \sum_j H_{ij} c_j = E_i c_i,
\end{equation}
where $H_{ij}=\langle\Phi_i|H|\Phi_j\rangle$. This is solved in matrix form, where standard diagonalization routines can be used to find the lowest eigenvalues and eigenvectors. The energy matrix of the CI method can be obtained as a projection of the exact Hamiltonian $H$ onto the CI subspace

\begin{equation}
    H^\mathrm{CI}=E_\mathrm{core}+\sum_{i>N_\mathrm{core}}h_i^\mathrm{CI}+\sum_{j>i>N_\mathrm{core}}V_{ij},
\end{equation}
where $E_\mathrm{core}$ is the energy of the frozen core, $N_\mathrm{core}$ is the number of core electrons, $h_i^\mathrm{CI}$ accounts for the kinetic energy of the valence electrons and their interaction with the central field, and $V_{ij}$ accounts for the valence-valence interaction. 

Defining $|J M\rangle$ and $|J^\prime M^\prime\rangle$ as many-electron states obtained from CI, with total angular momenta $J$ and $J^\prime$, and projections $M$ and $M^\prime$, a density matrix can be formed in terms of one-electron states $|nljm\rangle$

\begin{equation}
    \hat{\rho}=\rho_{nljm,n^\prime l^\prime j^\prime m^\prime}|nljm\rangle\langle n^\prime l^\prime j^\prime m^\prime|, 
\end{equation}
where

\begin{equation}
    \rho_{nljm,n^\prime l^\prime j^\prime m^\prime}=\langle J^\prime M^\prime|a_{n^\prime l^\prime j^\prime m^\prime}^\dagger a_{nljm}|JM\rangle. 
\end{equation}
Here, unprimed indices refer to the initial state, while primed indices refer to the final state. The many-electron matrix element can then be written as 

\begin{equation}
    \langle J^\prime M^\prime|T_q^L|JM\rangle=\mathrm{Tr}\,\rho_{nljm,n^\prime l^\prime j^\prime m^\prime}\langle n^\prime l^\prime j^\prime m^\prime|T_q^L|nljm\rangle,
\end{equation}
where the trace sums over all quantum numbers $(nljm)$ and $(n^\prime l^\prime j^\prime m^\prime)$, and $T_q^L$ is the spherical component of the tensor operator of rank $L$. Invoking the Wigner-Eckart theorem, one obtains the reduced matrix element

\begin{equation}
    \langle J^\prime \Vert T^L \Vert J\rangle = \mathrm{Tr}\,\rho_{nlj,n^\prime l^\prime j^\prime}^L \langle n^\prime l^\prime j^\prime\Vert T^L \Vert nlj\rangle, 
\end{equation}
where

\begin{multline}
    \rho_{nlj,n^\prime l^\prime j^\prime}^L = (-1)^{J^\prime -M^\prime}\left(
    \begin{array}{ccc}
        J^\prime & L & J \\ -M^\prime & q & M
    \end{array}\right)^{-1} \\
    \times \sum_{mm^\prime} (-1)^{j^\prime-m^\prime}\left(
    \begin{array}{ccc}
        j^\prime & L & j \\ -m^\prime & q & m
    \end{array}\right) \rho_{nljm,n^\prime l^\prime j^\prime m^\prime}. 
\end{multline}

\section{Overview of the package}\label{sec:overview}

In this section, we provide an overview of the core functionality of the pCI software package. 
The first task is the basis set construction, which is handled by the \texttt{hfd} and \texttt{bass} programs. 
After the basis set construction, a list of even or odd parity configurations to define the CI space is formed using the \texttt{add} program. 
The CI calculation can be run using the programs \texttt{pbasc} and \texttt{pconf} to obtain the desired energies and wave functions for each parity. 
Atomic properties such as $g$-factors, electric and magnetic multipole transition matrix elements, and hyperfine structure constants can then be computed using the \texttt{pdtm} program using the computed wave functions. 
Finally, valence polarizabilities can be obtained using the \texttt{pol} program using the computed matrix elements and wave functions.

\begin{figure*}[t]
    \centering
    \includegraphics[width=1\linewidth]{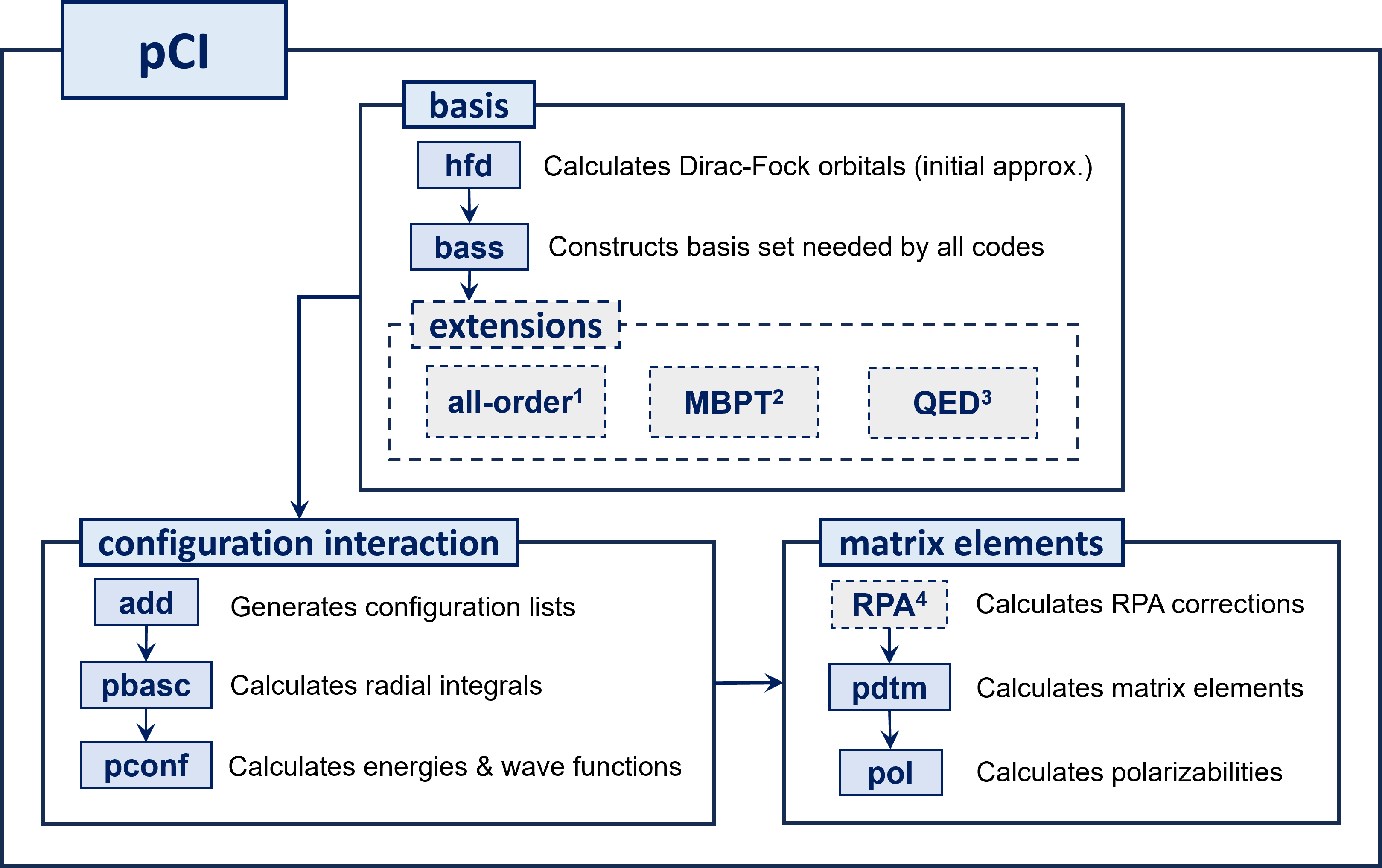}
    \caption{\label{fig:pci} Schematic of the pCI software package. The solid light blue boxes represent the core programs, while the dashed gray boxes represent optional modules. \\
    $^1$ A Fortran 77 version of the all-order package is included in the pCI software package, under the \texttt{lib/all-order} directory. An updated Fortran 90 version of the package will be released at a future date. \\
    $^2$ A Fortran 77 version of the MBPT package can be found in Ref.~\cite{KozPorSaf15}. \\
    $^3$ The QED package is included in the pCI software package, under the \texttt{lib/qed} directory. \\
    $^4$ The RPA package is included in the pCI software package, under the \texttt{lib/rpa} directory.}
\end{figure*}

\subsection{Installation}\label{sec:installation}
pCI has been developed and tested on Linux operating systems,
using the OpenMPI library with freely available Intel Fortran compilers. 
It has been designed to be run on high performance computing platforms.
Successful compilation and execution of the programs is not guaranteed on other operating systems, compilers or MPI implementations. 
Several programs also utilize Intel MKL's built-in LAPACK and ScaLAPACK routines to solve eigenproblems.

Installation instructions are kept updated in the \texttt{INSTALL.md} file in the root directory of the package. 
The software package is developed and maintained on GitHub (\url{https://github.com/ud-pci/pCI}), and
the documentation is hosted by Read the Docs (\url{https://pci.readthedocs.io}). 
Users can download the latest version of the pCI software package from the GitHub repository or clone the latest version using git: 

\begin{verbatim}
    git clone https://github.com/ud-pci/pCI.git
\end{verbatim}

After downloading the source files, the programs can be compiled using the CMake build tool and the \texttt{CMakeLists.txt} files in the root and \texttt{src} directories. Additional optimized and debugging builds can be done in a standard way as described in the \texttt{INSTALL.md} file.  

The \texttt{/lib} directory of the pCI distribution includes compatible Fortran 77 (F77) versions of a QED package~\cite{QEDMOD}, all-order package~\cite{Safronova08}, and RPA programs~\cite{KozPorSaf15}. There is also a Makefile that allows simple compilation of the respective programs via standard \texttt{make} execution. Running \texttt{make install} will install all F77 programs to the \texttt{/bin} directory, where the pCI programs are installed.

\section{Description of programs}\label{sec:programs}

In this section, we describe the main programs of the pCI software package in more detail. 

\begin{table*}[ht]
    \begin{threeparttable}
    \caption{\label{tab:pci_io_files} List of the pCI software package programs and their input and output files. The columns ``Program'' and ``Description'' lists the names of the programs and their main capability, respectively. The column ``Input (text)'' lists user-defined text input files. The column ``Input (binary)'' lists binary input files that are themselves constructed from a program (the \texttt{SGC.CON} file is a text file). The files given in parentheses are optional and used/created only in certain conditions. See Section~\ref{sec:pci_io} for more detail about the various input and output files. }
    \centering
    \begin{tabular}{cccccc}
        \hline \\[-3mm]
        Program & Description & Input (text) & Input (binary) & Output & Results \\[1mm]
        \hline \\[-3mm]
         \texttt{hfd}  & Solves HFD equations & \texttt{HFD.INP} & \texttt{(HFD.DAT)} & \texttt{HFD.DAT} & \texttt{HFD.RES} \\[2mm]
         \texttt{bass} & Constructs basis set & \texttt{BASS.INP} & \texttt{HFD.DAT} & \texttt{HFD.DAT} & \texttt{BASS.RES} \\[2mm]
         \texttt{add}  & Constructs configuration list & \texttt{ADD.INP} & & \texttt{CONF.INP} & \\[2mm]
         \texttt{pbasc} & Calculates radial integrals & \texttt{CONF.INP} & \texttt{HFD.DAT} & \texttt{CONF.DAT} & \texttt{BASC.RES} \\
         & & & & \texttt{CONF.INT} & \\
         & & & & \texttt{CONF.GNT} & \\[2mm]
         \texttt{pconf} & Performs CI calculation & \texttt{CONF.INP} & \texttt{CONF.DAT} & \texttt{CONF.DET} & \texttt{CONF.RES} \\
         & & \texttt{ci.in} & \texttt{CONF.INT} & \texttt{CONF.XIJ} & \texttt{CONF.ENG} \\
         & & & \texttt{CONF.GNT} & (\texttt{CONFp.JJJ}) & \texttt{CONF.LVL} \\
         & & & (\texttt{SGC.CON}) & (\texttt{CONFp.HIJ}) & \texttt{FINAL.RES} \\
         & & & (\texttt{SCRC.CON}) & & \texttt{LEVELS.RES} \\
         & & & & & \texttt{CONFSTR.RES} \\[2mm]
         \texttt{pdtm} & Calculates matrix elements & \texttt{CONF.INP} & \texttt{CONF.DAT} & \texttt{DTM.INT} & (\texttt{DM.RES}) \\
         & & \texttt{dtm.in} & \texttt{CONF.DET} & & or \\
         & & & \texttt{CONF.XIJ} & & (\texttt{TM.RES}) \\ 
         & & & (\texttt{CONF1.DET}) & & (\texttt{Operator.RES})$^{a}$ \\
         & & & (\texttt{CONF1.XIJ}) & &  \\[2mm]
         \texttt{pol} & Calculates polarizabilities & \texttt{CONF.INP} & \texttt{DTM.INT} & & \texttt{POL.RES} \\
         & & \texttt{pol.in} & \texttt{CONF.DAT} & & \texttt{POL\_E1.RES}\\
         & & & \texttt{CONF.HIJ}$^b$ & \\
         & & & \texttt{CONF.JJJ}$^b$ & \\
         & & & \texttt{CONF.DET} & \\
         & & & \texttt{CONF.XIJ} & \\
         & & & \texttt{CONF0.DET} & \\
         & & & \texttt{CONF0.XIJ} & \\[2mm]
         \hline
    \end{tabular}
    \begin{tablenotes}
       \item [$^a$] Separate files are generated for each specified operator (see Section~\ref{sec:dtm} for more detail).
       \item [$^b$] The \texttt{pconf} program creates the \texttt{CONFp.HIJ} and \texttt{CONFp.JJJ} files. These are processed by another program \texttt{sort}, which sorts them into the right format for \texttt{pol} (see Section~\ref{sec:pol} for more detail).
     \end{tablenotes}
    \end{threeparttable}
\end{table*}

\subsection{Input and output files}\label{sec:pci_io}
Before discussing the individual programs, we give brief descriptions of all the input and output files associated with the pCI software package. 

Each program requires an input file:
\begin{itemize}[noitemsep]
    \item \texttt{HFD.INP} - list of parameters for \texttt{hfd} program
    \item \texttt{BASS.INP} - list of parameters for \texttt{bass} program
    \item \texttt{ADD.INP} - list of parameters for \texttt{add} program
    \item \texttt{CONF.INP} - list of parameters for \texttt{pbasc}, \texttt{pconf}, \texttt{pdtm}, and \texttt{pol} programs
\end{itemize}

and outputs the following files:
\begin{itemize}[noitemsep]
    \item \texttt{hfd} and \texttt{bass}
    \begin{itemize}
        \item \texttt{HFD.DAT} - basis set radial orbitals $\phi_{nlj}$ and radial derivatives of the orbitals $\partial_r\phi_{nlj}$
        \item \texttt{HFD.RES} - contains results of \texttt{hfd} program
        \item \texttt{BASS.RES} - contains results of \texttt{bass} program
    \end{itemize}
    \item \texttt{pbasc}
    \begin{itemize}
        \item \texttt{CONF.DAT} - basis set radial orbitals $\phi_{nlj}$ and functions $\chi_{nlj}=h^\mathrm{r}_\mathrm{DF}\phi_{nlj}$, where $h^\mathrm{r}_\mathrm{DF}$ is the radial part of the DF operator
        \item \texttt{CONF.GNT} - relativistic Gaunt coefficients $G^k_q(fi)$~(Eq.~\ref{eq:gaunt})
        \item \texttt{CONF.INT} - one- and two-electron radial integrals
        \item \texttt{BASC.RES} - contains results of \texttt{pbasc} program
    \end{itemize}
    \item \texttt{pconf}
    \begin{itemize}
        \item \texttt{CONF.DET} - list of determinants
        \item \texttt{CONFp.HIJ} - list of matrix elements of the Hamiltonian (generated only if \texttt{Kw=1} in \texttt{ci.in})
        \item \texttt{CONFp.JJJ} - list of matrix elements of the operator $J^2$ (generated only if \texttt{Kw=1} in \texttt{ci.in})
        \item \texttt{CONF.XIJ} - calculated eigenvectors and eigenvalues
        \item \texttt{CONF.ENG} - tables of calculated quantum numbers $J_i$ and energy eigenvalues $E_i$ calculated each time \texttt{CONF.XIJ} is constructed during the Davidson procedure
        \item \texttt{CONF.LVL} - tables of top contributing configurations for each energy level calculated each time \texttt{CONF.XIJ} is constructed during the Davidson procedure
        \item \texttt{CONF.RES} - results of \texttt{pconf} program
        \item \texttt{FINAL.RES} - final table of quantum numbers $J_i$ and energy eigenvalues $E_i$ 
        \item \texttt{LEVELS.RES} - final table of top contributing configurations for each energy level 
        \item \texttt{CONFSTR.RES} - list of top contributing configurations along with their atomic term symbol for each energy level
    \end{itemize}
    \item \texttt{pdtm}
    \begin{itemize}
        \item \texttt{DTM.INT} - radial integrals for all one-electron operators included in \texttt{pdtm} program 
        \item \texttt{DTM.RES} - contains results of \texttt{pdtm} program
        \item \texttt{Operator.RES} - optional tables summarizing results of matrix element calculations for the specified operators
    \end{itemize}
    \item \texttt{pol}
    \begin{itemize}
        \item \texttt{POL.RES} - contains results of \texttt{pol} program
        \item \texttt{POL\_E1.RES} - final table of E1 polarizabilities
    \end{itemize}
\end{itemize}

\subsection{basis sets}\label{sec:basis}
The \texttt{hfd} and \texttt{bass} programs are responsible for the construction of the basis sets used in the CI calculations. \texttt{hfd} first solves the HFD equations to obtain an initial approximation, which \texttt{bass} can then use to construct the basis set.

\subsubsection{hfd}
The \texttt{hfd} program solves the restricted Hartree-Fock-Dirac (HFD) equations self-consistently under the central field approximation to find four-component Dirac-Fock (DF) orbitals and eigenvalues of the HFD Hamiltonian. The program provides the initial approximation, storing both one-electron basis radial orbitals
\begin{equation}
    \phi_{nlj} = r \left(
    \begin{array}{c}
         f_{nlj}  \\
         -g_{nlj} 
    \end{array}\right),
\end{equation}
as well as the radial derivatives of the orbitals, $\partial_r\phi_{nlj}$. 
A more detailed description of this program is given in Refs.~\cite{BDT77,KozPorSaf15}. 

\subsubsection{bass}
The \texttt{bass} program constructs the basis set starting from the HFD orbitals for the core and valence orbitals, formed by the \texttt{hfd} program. Then virtual orbitals are added to account for correlations and are constructed from either (1) previously constructed HFD orbitals or (2) B-splines. 

In the first case, virtual orbitals are formed using a recurrent procedure described in~\cite{KozPorFla96,KozPorSaf15}.  The lowest virtual orbitals can be constructed from the HFD orbitals. 
The large component of the radial Dirac bispinor, $f_{n'l'j'}$, is obtained from a function $f_{nlj}$ constructed previously by multiplying it by $r^{l' - l}\, \sin(kr)$. Here $l'$ and $l$ are the orbital quantum numbers of the new and old orbitals ($l' \geq l$) and the coefficient $k$ is determined by the properties of the radial grid. The small component $g_{n'l'j'}$ is found from the kinetic balance condition:
\begin{equation}
\label{kbal}
g_{n'l'j'} =\frac{\boldsymbol{\sigma} \bf p}{2mc} f_{n'l'j'} ,
\end{equation}
where $\boldsymbol{\sigma}$ are the Pauli matrices, ${\bf p}$ and $m$ are the electron momentum and mass, and $c$ is the speed of light.
The newly constructed functions are then orthonormalized to the functions of the same symmetry.

Another option is to construct large components of the orbitals from B-splines. Small components are still formed with the kinetic balance method. A more detailed description of this program is given in Ref.~\cite{KozPorSaf15}.

\subsection{configuration interaction}\label{sec:ci}
The CI method is implemented collectively by three programs: \texttt{add}, \texttt{pbasc}, and \texttt{pconf}. We begin with \texttt{add}, which constructs a CI space defined by a list of relativistic configurations generated by allowing excitations to any orbital in the basis set constructed in the previous section. 
The \texttt{pbasc} program acts as a precursor to \texttt{pconf}, calculating and storing the radial integrals required to construct the Hamiltonian matrix in the CI space.
The \texttt{pconf} program constructs the Hamiltonian matrix then diagonalizes it to find low-lying energies and eigenvectors. 

\subsubsection{add}
The \texttt{add} program constructs a list of configurations to define the CI space by exciting electrons from a set of reference configurations to a set of active nonrelativistic shells. It takes in the input file \texttt{ADD.INP}, which specifies the reference configurations, active nonrelativistic shells, and minimum and maximum occupation numbers for each shell. As output, it writes the file \texttt{CONF.INP}, which includes a list of user-definable parameters and the list of configurations constructed by exciting electrons from a list of reference configurations to the orbitals in the basis set. A sample \texttt{ADD.INP} file is displayed in Figure~\ref{fig:add_even} for constructing a list of even-parity configurations for Fe$^{16+}$. A description of each parameter is defined after the ``\#'' symbol. 
A more detailed description of this program is given in Ref.~\cite{KozPorSaf15}.

\subsubsection{pbasc}
After creating the configuration list, we pre-calculate the one- and two-electron radial integrals using the parallel program \texttt{pbasc}. These integrals are used by the proceeding parallel program \texttt{pconf} to form the Hamiltonian matrix in the CI space. The one-electron radial integrals correspond to the DF potential of the core, and the two-electron radial integrals account for the Coulomb, and optionally Breit interaction between the valence electrons. The matrix elements of the Coulomb interaction for the multipolarity $k$ can be written as

\begin{equation}
    \langle c d|V_q^k|a b\rangle \equiv G_q^k(ca) G_q^k(bd) R_{abcd}^k,  
\end{equation}
where $G_q^k(fi)$ are angular factors, or relativistic Gaunt coefficients, given by
\begin{multline}
\label{eq:gaunt}
    G_q^k(fi)=(-1)^{m_f+1/2}\delta_p\sqrt{(2j_i+1)(2j_f+1)} \\ \times
        \begin{pmatrix} 
         j_f & j_i & k \\  
        -m_f & m_i & q
        \end{pmatrix}
        \begin{pmatrix} 
        j_f & j_i & k \\  
        1/2 & -1/2 & 0
        \end{pmatrix},
\end{multline}
$R_{abcd}^k$ are the relativistic Coulomb radial integrals, and $\delta_p$ accounts for the parity selection rule

\begin{equation}
    \delta_p=\xi(l_i+l_f+k), \hspace{0.2in}\xi(n)=\Bigg\{
    \begin{matrix}
    1 & \mathrm{if}\;n\;\mathrm{is\;even}, \\ 
    0 & \mathrm{if}\;n\;\mathrm{is\;odd}.
    \end{matrix} 
\end{equation}

The Breit interaction has the same form as the Coulomb interaction, but without the parity selection rule. 

\texttt{pbasc} reads the files \texttt{HFD.DAT} and \texttt{CONF.INP} to determine which radial integrals are needed. These integrals are calculated and written to the file \texttt{CONF.INT}. The relativistic Gaunt coefficients are written to the file \texttt{CONF.GNT}, and the file \texttt{CONF.DAT} is also formed, storing the basis radial orbitals $\phi_{nlj}$, as well as functions $\chi_{nlj} = h^\mathrm{CI}_i\phi_{nlj}$. 

\subsubsection{pconf}
The parallel program \texttt{pconf} performs the configuration interaction method in the CI space defined by the list of configurations contained in the \texttt{CONF.INP} input file created by the \texttt{add} program. It takes the \texttt{CONF.DAT}, \texttt{CONF.GNT}, \texttt{CONF.INT}, and \texttt{CONF.INP} files as input. In addition, \texttt{pconf} can also take in the \texttt{SGC.CON} and \texttt{SCRC.CON} files, which contain one- and two-electron effective radial integrals, respectively. These files are created from supplementary programs from the serial CI-MBPT and to-be-released all-order code packages to include additional MBPT and all-order corrections. The inclusion of QED corrections also utilizes these optional files.
To run the program, we must first create the input file \texttt{ci.in}, which contains a list of key-value pairs defining the CI computation:
\begin{verbatim}
    # ci.in
    Kl = (0, 1, 2, 3)
    Ksig = (0, 1, 2)
    Kdsig = (0, 1)
    Kw = (0, 1)
    kLSJ = (0, 1)
\end{verbatim}

The value of \texttt{Kl} can take the following values:
\begin{itemize}[noitemsep]
    \item \texttt{0} - start a new CI calculation
    \item \texttt{1} - continue the CI calculation (requires \texttt{Kw=1})
    \item \texttt{2} - start a new CI calculation, including corrections from MBPT/all-order/QED, etc.
    \item \texttt{3} - continue a CI calculation, including more configurations (requires \texttt{Kw=1})
\end{itemize}
Note that for \texttt{Kl=1,3} to work, the file \texttt{CONFp.HIJ} must have been successfully written with \texttt{Kw=1} in the original run. \texttt{CONFp.HIJ} stores the previous Hamiltonian matrix, which must be read to continue a CI calculation. However, this file can be as big as 1~TB or more for large systems, so it should not be used unless the user has exceptional available computational resources. 

The value of \texttt{Ksig} can take the following values:
\begin{itemize}[noitemsep]
    \item \texttt{0} - pure CI
    \item \texttt{1} - include one-electron corrections
    \item \texttt{2} - include one- and two-electron corrections
\end{itemize}

The value of \texttt{Kdsig} can take the following values:
\begin{itemize}[noitemsep]
    \item \texttt{0} - automatic approximation of the energy dependence of the operator $\Sigma(E)$
    \item \texttt{1} - manually specify the energy $E_\textrm{val}$ to treat the energy dependence of $\Sigma(E)$
\end{itemize}

The value of \texttt{Kw} can take the following values:
\begin{itemize}[noitemsep]
    \item \texttt{0} - do not write the Hamiltonian matrix to file \texttt{CONFp.HIJ}
    \item \texttt{1} - write the Hamiltonian matrix to file \texttt{CONFp.HIJ}
\end{itemize}

The value of \texttt{kLSJ} can take the following values:
\begin{itemize}[noitemsep]
    \item \texttt{0} - do not calculate $\langle S^2 \rangle$, $\langle L^2 \rangle$, or form approximate terms for each energy level
    \item \texttt{1} - calculate $\langle S^2 \rangle$, $\langle L^2 \rangle$, and form approximate atomic term symbols for each energy level
\end{itemize}

After reading the keys from \texttt{ci.in}, the \texttt{pconf} program reads the general parameters and the list of configurations from the \texttt{CONF.INP} file. Next, the information about the basis set is read from \texttt{CONF.DAT}, relativistic Gaunt coefficients are read from \texttt{CONF.GNT}, radial integrals are read from \texttt{CONF.INT}, and optionally, effective radial integrals are read from \texttt{SGC.CON} and \texttt{SCRC.CON}. 

Having read all required input files, the \texttt{pconf} program forms a list of determinants from the list of relativistic configurations and writes them to the file \texttt{CONF.DET}. With the list of determinants, it forms the Hamiltonian matrix and then the matrix of the operator $J^2$. The construction of the Hamiltonian matrix is the most time-consuming part of the \texttt{pconf} program. A detailed description of the Hamiltonian matrix construction and its MPI implementation can be found in Ref.~\cite{sym13040621}.

After the matrices are constructed, the \texttt{pconf} program enters the Davidson iterative procedure, where the Hamiltonian is diagonalized to obtain a specified number of low-lying energy eigenvalues and eigenvectors. The progress of the Davidson iterative procedure is written in the \texttt{CONF.PRG} file at each iteration. At selected intervals before the final iteration, the eigenvalues and eigenvectors are saved to \texttt{CONF.XIJ}. Each time \texttt{CONF.XIJ} is written on disk, a table of the energy levels is appended to the file \texttt{CONF.ENG}, and tables of the top contributing configurations for each level are appended to \texttt{CONF.LVL}. 

Once the Davidson iterative procedure has converged, the final eigenvalues and eigenvectors are saved once again to \texttt{CONF.XIJ}, the final energy table is saved to \texttt{FINAL.RES}, the final list of the top contributing configurations for each level is saved to \texttt{LEVELS.RES}, and the configurations along with their atomic term symbol is written to \texttt{CONFSTR.RES}. 

If polarizability calculations are required, the Hamiltonian matrix elements will have to be written to the file \texttt{CONFp.HIJ} by setting the key \texttt{Kw=1} in \texttt{ci.in}. Note that depending on the size of the Hamiltonian, this file could take up hundreds of GB to over 1 TB. By default, \texttt{Kw=0} is set to not write the Hamiltonian to disk. As a reference, polarizability calculations have typically been done with Hamiltonian sizes of up to 4 million determinants. 

\subsection{matrix elements}

\subsubsection{pdtm}\label{sec:dtm}
The \texttt{pdtm} program calculates the matrix elements of one-electron operators between many-electron states, under the density (or transition) matrix formalism. This formalism allows us to express the matrix elements between many-electron states via one-electron matrix elements. The \texttt{pdtm} program forms these reduced density (or transition) matrices and calculates the reduced matrix elements. The following quantities can be calculated from this program:  

\begin{itemize}[noitemsep]
    \item electron $g$-factors  
    \item magnetic dipole and electric quadrupole hyperfine structure constants $A$ and $B$  
    \item electric $Ek$ and magnetic $Mk$ multipole transition amplitudes, where $k = 1,2,3$ corresponds to the dipole, quadrupole, and octupole transitions  
    \item nuclear spin independent parity nonconserving (PNC) amplitude  
    \item amplitude of the electron interaction with the P-odd nuclear anapole moment (AM) 
    \item P, T-odd interaction of the electron electric dipole moment  
    \item amplitude of the electron interaction with the P, T-odd nuclear magnetic quadrupole moment
\end{itemize}

The \texttt{pdtm} program takes in the input file \texttt{dtm.in}, which defines the parameters of the matrix element computation:
\begin{verbatim}
    Mode = (DM, TM, Init)
    Levels = level_range (level_range_1)
    Operators = (E1, M2, ...) (optional)
\end{verbatim}
The value of \texttt{Mode} can take the following values:
\begin{itemize}[noitemsep]
    \item \texttt{DM} - form density matrix and calculate diagonal matrix elements
    \item \texttt{TM} - form transition matrix and calculate non-diagonal matrix elements
    \item \texttt{Init} - create the file \texttt{DTM.INT}
\end{itemize}

The value of \texttt{Levels} takes in a range of levels depending on the whether \texttt{Mode} was specified as \texttt{DM} or \texttt{TM}:
\begin{itemize}[noitemsep]
    \item \texttt{DM}
    \begin{itemize}
        \item \texttt{initial\_level} \texttt{final\_level}
    \end{itemize}
    \item \texttt{TM}
    \begin{itemize}
        \item \texttt{initial\_level} \texttt{final\_level} \texttt{initial\_level\_1} \texttt{final\_level\_1}
    \end{itemize}
\end{itemize}
This can be left empty for the case of \texttt{Mode = Init}, where only the \texttt{DTM.INT} file needs to be written. The \texttt{initial\_level} and \texttt{final\_level} correspond to the levels written to \texttt{CONF.XIJ}, and those with ``\texttt{\_1}'' correspond to levels in \texttt{CONF1.XIJ}. 

The value of \texttt{Operators} takes in a list of operators to create additional files summarizing the results in a table:
\begin{itemize}[noitemsep]
    \item \texttt{DM}
    \begin{itemize}
        \item \texttt{GF}, \texttt{A\_hf}, \texttt{B\_hf}
    \end{itemize}
    \item \texttt{TM}
    \begin{itemize}
        \item \texttt{E1}, \texttt{E2}, \texttt{E3}, \texttt{M1}, \texttt{M2}, \texttt{M3}, \texttt{EDM}, \texttt{PNC}, \texttt{AM}, \texttt{MQM}
    \end{itemize}
\end{itemize}
Note that this key is optional and unnecessary for the core functionality of \texttt{pdtm}. A table with the results is written to the files \texttt{DM.RES} or \texttt{TM.RES}, regardless. 

This program begins by reading the file \texttt{CONF.INP} for the system parameters and the list of configurations. Then, the basis radial orbitals are read from the file \texttt{CONF.DAT}, and radial integrals for all operators are calculated and written to the file \texttt{DTM.INT}. If this file already exists, \texttt{pdtm} uses it and does not recalculate the radial integrals. If \texttt{Mode = Init}, the file \texttt{DTM.INT} will be reconstructed regardless of whether the file already exists. 

For diagonal matrix elements, the list of determinants and eigenvectors corresponding to the state of interest is read from the files \texttt{CONF.DET} and \texttt{CONF.XIJ}, respectively. For the non-diagonal matrix elements, the initial state is read from the file \texttt{CONF.DET} and \texttt{CONF.XIJ}, and the final state is read from the files \texttt{CONF1.DET} and \texttt{CONF1.XIJ}. The results of the diagonal and non-diagonal matrix elements are written to the files \texttt{DM.RES} and \texttt{TM.RES}, respectively. 

\subsubsection{pol}\label{sec:pol}
The \texttt{pol} program calculates the dc and ac polarizabilities of the specified atomic states. The expression for electric-dipole ac polarizability at the frequency $\omega$ of the state $|JM\rangle$ can be written (in a.u.) as a sum over unperturbed intermediate states $n$,
%--------------------------------------------------------------------
\begin{equation}
\alpha(\omega) = 2 \sum_n \frac{(E_n-E) |\langle JM|D_z|n\rangle|^2}{(E_n-E)^2 - \omega^2} ,
\label{alpha}
\end{equation}
%--------------------------------------------------------------------
where $\bf D$ is an electric dipole moment operator and $E$ and $E_n$ are the energies of the initial and intermediate states, respectively.

To find $\alpha$, we can rewrite Eq.~(\ref{alpha}) as
%--------------------------------------------------------------------
\begin{eqnarray}
\alpha(\omega) &=& \sum_n \langle JM|D_z|n\rangle \,\langle n|D_z| JM\rangle \nonumber \\
       &\times& \left[ \frac{1}{E_n-E+\omega} + \frac{1}{E_n-E-\omega} \right] .  
\label{alpha1}
\end{eqnarray}
%--------------------------------------------------------------------

Then we use the Sternheimer~\cite{Ste50} or Dalgarno-Lewis~\cite{DalLew55} method and solve the inhomogeneous equations
%------------------------------------------------------------------
\begin{eqnarray}
(H - E \pm \omega)\, |\delta \phi_{\pm} \rangle = D_z\, |JM \rangle,
\label{inhom}
\end{eqnarray}
%------------------------------------------------------------------
to find $|\delta \phi_{\pm} \rangle$, 
%------------------------------------------------------------------
\begin{eqnarray}
|\delta \phi_{\pm} \rangle &=& \frac{1}{H - E \pm \omega} D_z |JM \rangle \nonumber \\
&=& \sum_n \frac{1}{H - E \pm \omega}|n\rangle \langle n |D_z |JM \rangle
\label{phi}
\end{eqnarray}
%------------------------------------------------------------------
where $H$ is the Hamiltonian and we used the closure relation 
$\sum_n | n \rangle \langle n | = 1$.
After that, the polarizability can be found as the sum of two matrix elements
\begin{eqnarray}
\alpha(\omega) =  \langle JM| D_z |\delta \phi_{+} \rangle + 
                  \langle JM| D_z |\delta \phi_{-} \rangle .
\label{alpha2}
\end{eqnarray}

If electrons in an atomic system are divided into valence and core electrons, the polarizability can be divided accordingly as
\[
\alpha \equiv \alpha_v + \alpha_c, %+ \alpha_{vc}  
\]
where $\alpha_v$ and $\alpha_c$ are the valence and core contributions.
 %and $\alpha_{vc}$ is a part of the core contribution that appears due to a possible excitation of a core electron to the occupied valence state. The Pauli principle forbids this, so this contribution should be subtracted from $\alpha_c$.

The \texttt{pol} program calculates only the valence polarizability $\alpha_v$. The core polarizability needs to be computed separately with a different program. 

Disregarding the vector polarizability, we can present the expression for $\alpha(\omega)$ as the sum of the scalar and tensor parts,
\begin{equation}
\alpha(\omega) = \alpha_0 + \alpha_2 \, \frac{3M^2-J(J+1)}{J(2J-1)} .
\label{al_st}
\end{equation}
The \texttt{pol} program gives both the scalar and tensor polarizabilities if the latter is not zero.
%, but not the vector polarizability. 

This program requires several input files from previously run \texttt{pconf} and \texttt{pdtm} programs, including \texttt{CONF.DET} and \texttt{CONF.XIJ} of the parity of the level of interest (renamed to \texttt{CONF0.DET} and \texttt{CONF0.XIJ}), \texttt{CONF.INP}, \texttt{CONF.XIJ}, \texttt{CONF.HIJ}, and \texttt{CONF.JJJ} of the opposite parity, and the file \texttt{DTM.INT} from \texttt{pdtm}.

Note that the \texttt{pconf} program outputs \texttt{CONFp.HIJ} and \texttt{CONFp.JJJ} files, but not the \texttt{CONF.HIJ} and \texttt{CONF.JJJ} files. The only difference between these files is that the former are not sorted and require the additional \texttt{sort} program to process them in the latter. 

The \texttt{pol} program requires a list of key-value parameters in a file \texttt{pol.in}:
\begin{verbatim}    
    Mode = (0, 1)
    Method = (0, 1, 2)
    Level = (energy level index)
    Ranges = (list of ranges of wavelengths)
    IP1 = (dimension of initial matrix)
\end{verbatim}

The value of \texttt{Mode} can take the following values:
\begin{itemize}
    \item 0 - start a new calculation
    \item 1 - continue the calculation
\end{itemize} 

The value of \texttt{Method} can take the following values:
\begin{itemize}
    \item 0 - invert the matrix and iterate if diverged
    \item 1 - only invert the matrix 
    \item 2 - modified iteration procedure where computation restarts after every 2 iterations while retaining vectors (used in cases where \texttt{Method=0} diverges)
\end{itemize}

The value of \texttt{Level} is the ordinal number of the vector in the \texttt{CONF0.XIJ} file corresponding to the energy level for which the user wants to calculate the polarizability. 
The \texttt{Ranges} field takes in a list of wavelength ranges with step size in the format (\texttt{initial\_wavelength final\_wavelength step\_size}), separated by commas. 
For example, the range ``\texttt{0 0 0}'' corresponds to calculations of dc polarizabilities, while ``\texttt{500 505 0.5}'' tells \texttt{pol} to calculate ac polarizabilities from $\lambda=500$ nm to $\lambda=505$ nm in steps of \texttt{0.5} nm. 
The value of \texttt{IP1} sets the dimension of the initial approximation of the matrix (by default, this is set to \texttt{IP1=15000}).

An example of running the program \texttt{pol} is given in Sec.~\ref{sec:sr}.

\section{Scalability and efficiency of parallel codes}\label{sec:scalability}

Large-scale CI calculations are typically run on large computing clusters with large amounts of processors and memory. The parallel programs contained in the pCI software package utilize MPI to massively parallelize their most time-consuming algorithms. MPI allows for scalable parallelism over many distributed computing nodes in modern parallel computing architectures. Details of the new algorithms used in the pCI package over the previous CI-MBPT package can be found in Ref.~\cite{sym13040621}. In summary, the parallel codes utilize dynamic memory allocations to optimize the memory footprint and dynamic load-balancing to optimize heavy workload calculations.   

To demonstrate the scalability and efficiency of the parallel codes, we conducted a scalability test using up to 32 nodes and 2048 computing cores on the University of Delaware DARWIN computing cluster. As a test case, we used a prior 30-electron CI calculation for low-lying Ir$^{17+}$ energy levels. Details of this calculation can be found in Ref.~\cite{Cheung2020}. We used a short $8spdfg$ basis set and included 24\,895 relativistic configurations (17\,431\,323 determinants) in the CI space. The Hamiltonian matrix contained about $27.76\times10^9$ nonzero matrix elements, which were stored in approximately 413.6 GiB of memory. 

The results of the scalability test are shown in Table~\ref{tab:scalability}. They are obtained from timing the main code sections in the program \texttt{pconf}: the construction of the Hamiltonian and its diagonalization. We find a perfect linear speedup with the number of nodes and cores for the construction of the Hamiltonian matrix. However, the matrix diagonalization does not perform as well because the Davidson procedure is iterative and has much more serial overhead. When many more cores take part in the calculation, the communication overhead also becomes considerable and contributes to the lackluster speedup. However, despite the poor parallelism of the Davidson diagonalization procedure, the total speedup drops to around 90\% when running with 8 nodes and to around 80\% with 16 nodes. We find similar scalability in the other parallel programs \texttt{pbasc} and \texttt{pdtm}. 

Large-scale runs are typically done with less than 16 nodes due to prohibitive memory requirements caused by the size of the CI space. Note that scalability and efficiency depend highly on the number of Davidson iterations required to obtain the convergence of the energy levels. The construction of the Hamiltonian matrix does not affect the scalability as much, due to the implemented dynamic load-balancing scheme~\cite{sym13040621}. We find that calculations where matrix construction dominates are more scalable than those where diagonalization dominates. Further work on the implemented matrix diagonalization scheme has to be done to achieve a perfect linear speedup for higher node counts for diagonalization-dominated calculations. Matrix-construction-dominated calculations are expected to perform well for a much higher number of cores but need to be tested on larger computational resource centers.    
 
\begin{table*}[ht]
    \centering
    \caption{Results of a scalability test done using a 30-electron CI calculation using an $8spdfg$ basis set for the energies of Ir$^{17+}$~\cite{Cheung2020}. The number of nodes, cores and the amount of memory per core (in GiB) is displayed in the column ``Computational Resources''. The execution time for the construction of the Hamiltonian matrix (Construct), the Davidson iterative procedure (Davidson), and the total execution time (Total) are listed in the column ``Time''. The speedup achieved for each parallel run is displayed in the column ``Speedup'', and is calculated from the base 1-node, 64-core run. Note that this calculation took about 2 weeks of execution time using the serial code~\cite{KozPorSaf15, Cheung2020}.}
    \begin{tabular}{|c|c|c|c|c|c|c|c|c|}
        \hline
        \multicolumn{3}{|c|}{Computational Resources} & 
        \multicolumn{3}{|c|}{Time} &
        \multicolumn{3}{|c|}{Speedup} \\\hline
        nodes & cores & mem\_per\_core (GiB) & Construct & Davidson & Total & Construct & Davidson & Total \\\hline
          1 &   64 & 9.1 &  4h 32m & 28m 56s &  5h  3m & - & - & -  \\
          2 &  128 & 5.5 &  2h 15m & 16m 27s &  2h 32m &  2.02 & 1.76 & 1.98 \\
          4 &  256 & 3.7 &  1h  7m &  9m 57s &  1h 18m &  4.05 & 2.91 & 3.84 \\
          8 &  512 & 2.8 & 33m 27s &  6m 18s & 41m 31s &  8.13 & 4.59 & 7.30 \\
         16 & 1024 & 2.4 & 16m 41s &  4m 33s & 23m  3s & 16.29 & 6.36 & 13.14 \\
         32 & 2048 & 2.1 &  8m 21s &  4m  2s & 14m 57s & 32.55 & 7.17 & 20.26 \\\hline
    \end{tabular}
    \label{tab:scalability}
\end{table*}

\section{Calculation of Fe\(^{16+}\) energies and 3C/3D ratio}\label{sec:fe_xvii}
In this section, we showcase a sample workflow of the pCI software package, calculating the energies and 3C/3D oscillator strength ratios in Fe$^{16+}$~\cite{2022Fe16+,2024Fe16+}. 
This method described here has also been used for calculations for other highly charged ions, such as Ni$^{18+}$~\cite{2024Ni18+}. 
Additional examples are available on the documentation website. 
By the end of this example, we will have our results organized in the following sample file directory:
\begin{verbatim}
    /Fe16+
        /basis  - contains basis set files
        /dtm    - contains matrix element files
        /even   - contains even-parity CI files
        /odd    - contains odd-parity CI files
\end{verbatim}

\subsection{Construction of basis set}
We begin the construction of the basis set for Fe$^{16+}$ by first constructing the core and valence orbitals from solutions of the HFD equations using the \texttt{hfd} program, and then running the \texttt{bass} program to form virtual orbitals to account for correlations. Following Ref~\cite{2022Fe16+}, we construct a $24spdfg$ basis set, where the designation $24spdfg$ means that all orbitals up to $n=24$ are included for the $spdfg$ partial waves. Here, the $1s$, $2s$, $3s$, $3p$, $4s$, $4p$, $4d$, $4f$, and $5g$ orbitals are constructed as DF orbitals, while all others in the basis set are constructed in the usual way using \texttt{bass} following Refs.~\cite{KozPorFla96,KozPorSaf15}. A detailed description of the basis set construction can be found in the Fe$^{16+}$ example page on the documentation website: \url{https://pci.readthedocs.io/en/latest/examples/hci.html}. 
This program outputs the final basis set to the \texttt{HFD.DAT} file, which will be used for all the following programs. At this point, the \texttt{HFD.DAT} file should be copied or symlinked to the \texttt{dtm}, \texttt{even}, and \texttt{odd} directories.

\subsection{Calculation of energy levels}
Before we can run the CI calculations, we must first construct an even-parity and an odd-parity list of configurations that define the CI space for the \texttt{pconf} program. This is done using the \texttt{add} program with the \texttt{ADD.INP} input file for each parity, in their respective \texttt{/even} and \texttt{/odd} directories. Figure~\ref{fig:add_even} displays a sample even-parity \texttt{ADD.INP} file.
\begin{figure*}
    \caption{\label{fig:add_even} Contents of \texttt{ADD.INP} for even-parity configurations of Fe$^{16+}$. Parameters of the \texttt{CONF.INP} input file are defined in the comments to the right. Refer to Ref.~\cite{KozPorSaf15} for more details on the parameters.}
    \begin{small}
    \begin{verbatim}
                      # ADD.INP
                      Ncor=  2                  # number of reference configurations
                      NsvNR 49                  # number of allowed orbitals
                      mult=  2                  # multiplicity of excitations
                       NE =  8                  # number of valence electrons
                      
                      L:   2s2   2p6            # reference configuration #1
                      L:   2s2   2p5   3p1      # reference configuration #2
                      
                         2s  0  2   2p  0  6   3s  0  2   3p  0  6   3d  0  6   4s  0  2
                         4p  0  6   4d  0  6   4f  0  6   5g  0  6   5s  0  2   5p  0  6
                         5d  0  6   5f  0  6   6s  0  2   6p  0  6   6d  0  6   6f  0  6
                         6g  0  6   7s  0  2   7p  0  6   7d  0  6   7f  0  6   7g  0  6
                         8s  0  2   8p  0  2   8d  0  2   8f  0  2   8g  0  2   9s  0  2
                         9p  0  2   9d  0  2   9f  0  2   9g  0  2  10s  0  2  10p  0  2
                        10d  0  2  10f  0  2  10g  0  2  11s  0  2  11p  0  2  11d  0  2
                        11f  0  2  11g  0  2  12s  0  2  12p  0  2  12d  0  2  12f  0  2
                        12g  0  2  13s  0  2  13p  0  2  13d  0  2  13f  0  2  13g  0  2
                        14s  0  2  14p  0  2  14d  0  2  14f  0  2  14g  0  2  15s  0  2
                        15p  0  2  15d  0  2  15f  0  2  15g  0  2  16s  0  2  16p  0  2
                        16d  0  2  16f  0  2  16g  0  2  17s  0  2  17p  0  2  17d  0  2
                        17f  0  2  17g  0  2  18s  0  2  18p  0  2  18d  0  2  18f  0  2
                        18g  0  2  19s  0  2  19p  0  2  19d  0  2  19f  0  2  19g  0  2
                        20s  0  2  20p  0  2  20d  0  2  20f  0  2  20g  0  2  21s  0  2
                        21p  0  2  21d  0  2  21f  0  2  21g  0  2  22s  0  2  22p  0  2
                        22d  0  2  22f  0  2  22g  0  2  23s  0  2  23p  0  2  23d  0  2
                        23f  0  2  23g  0  2  24s  0  2  24p  0  2  24d  0  2  24f  0  2
                        24g  0  2    
                      >>>>>>>>>>>>> Head of the file CONF.INP >>>>>>>>>>>>>>>>>>>>>>>>
                        Fe16+_even                                                            
                        Z = 26.0        # atomic number
                       Am = 56.0        # atomic mass
                        J = 0           # total angular momentum
                       Jm = 0           # total angular momentum projection
                       Nso= 1           # number of core shells
                       Nc = 10          # number of configurations (placeholder)
                       Kv = 4           # key for diagonalization
                       Nlv= 5           # number of energy levels to compute
                       Ne = 8           # number of valence electrons
                       Kl4= 1           # key for initial approximation of eigenvectors
                       Nc4=999          # number of conf-s for initial approximation
                      Crt4= 0.0001      # convergence criterion for Davidson procedure
                      kout= 0           # key defining amount of output detail
                      Ncpt= 0           # number of conf-s to include in PT space
                      Cut0= 0.0001      # cutoff criteron for PT configurations
                      N_it= 50          # number of Davidson iterations
                      Kbrt= 2           # key for Breit interaction
                           0.1002       # list of core shells
                      ==================================================================
    \end{verbatim}
    \end{small}
\end{figure*}

Here, we keep the $1s$ shell closed and allow all single and double excitations from the 8 valence electrons from the 2 reference configurations $2s^22p^6$ and $2s^22p^53p$, to the 109 orbitals listed, up to $24spdfg$. From here on, we omit $1s^2$ from configuration designations for brevity. An equivalent odd-parity \texttt{ADD.INP} file would have the same content, with the reference configurations $2s^2 2p^5 3s$, $2s^2 2p^5 3d$ and $2s 2p^6 3p$, and the value of \texttt{Ncor} set to 3. Also note that the 3C and 3D lines are both $J=1 \to 0$ transitions, specifically transitions from the 4th and 5th lowest odd-parity $J=1$ states to the ground $J=0$ state. To minimize the use of computational resources, we can specify \texttt{J=1}, \texttt{Jm=1}, \texttt{Nlv=5}, and \texttt{Kv=3} to only compute the 5 lowest energy levels with $J=1$. These changes can either be made in the \texttt{CONF.INP} file, or at the bottom of the \texttt{ADD.INP} file. Alternatively, if enough computational resources are available, one can allow \texttt{J=0}, \texttt{Jm=0}, \texttt{Nlv=16}, and \texttt{Kv=4}, to calculate the lowest 16 energy levels with any allowed value of $J$. 
The next step is to run the sequence of programs \texttt{pbasc} and \texttt{pconf} in the respective even- and odd-parity directories. These are both parallel programs, so they have to be executed using the commands
\begin{verbatim}
    mpirun -n <nprocs> <executable>
\end{verbatim}
where \texttt{<nprocs>} specifies the number of MPI processes to use, and \texttt{<executable>} is the program, \texttt{pbasc} or \texttt{pconf}, to run. The \texttt{pconf} program outputs the basis set of determinants to the \texttt{CONF.DET} file and the wave functions to the \texttt{CONF.XIJ} file, which will be copied to the \texttt{/dtm} directory to be used for matrix element calculations using the \texttt{pdtm} program. 

The final energies are tabulated in the files \texttt{CONF.RES} and \texttt{FINAL.RES}. We present the results of the \texttt{pconf} program for the first 5 odd-parity levels with $J=1$ in Table~\ref{fe_table}. Our results are compared with experimental values from Ref.~\cite{2024Fe16+} and Ref.~\cite{AK}. Note that the sample calculation is done with a very small basis set of $12spdfg$. Higher accuracy at the level of 0.001\% as in Ref.~\cite{2024Fe16+} can be attained by reaching basis set convergence by re-running the computations with larger basis sets until energy differences are small. Other corrections include increasing the number of reference configurations, adding triple excitations, as well as including QED contributions~\cite{2024Fe16+}. 

\begin{table*}
\centering
\caption{\label{fe_table} Odd-parity energies of Fe$^{16+}$ calculated with a small $12spdfg$ basis set for the first 5 levels with $J=1$. Energies are calculated from the even-parity $2s^2 2p^6$ ground state. The results are compared with experiment from Ref.~\cite{2024Fe16+} and Ref.~\cite{AK}. All energies are given in cm$^{-1}$. }
\begin{tabular}{lcccccc}
\hline
\multicolumn{2}{c}{Configuration}&
\multicolumn{1}{c}{$12spdfg$}&
\multicolumn{1}{c}{Expt.~\cite{2024Fe16+}}&
\multicolumn{1}{c}{Diff.~\cite{2024Fe16+}} &
\multicolumn{1}{c}{Expt.~\cite{AK}}&
\multicolumn{1}{c}{Diff.~\cite{AK}} \\
\hline \\ [-0.8pc]
$2s^2 2p^5 3s $ &$^3P_1$ & 5860927 & 5864239 & 0.05\% & 5864502 & 0.05\% \\
$2s^2 2p^5 3s $ &$^1P_1$ & 5956909 & 5960977 & 0.06\% & 5960742 & 0.06\% \\
$2s^2 2p^5 3d $ &$^3P_1$ & 6468748 &     -   &    -   & 6471640 & 0.03\% \\
$2s^2 2p^5 3d $ &$^3D_1$ & 6550091 & 6552585 & 0.03\% & 6552503 & 0.03\% \\
$2s^2 2p^5 3d $ &$^1P_1$ & 6658398 & 6661091 & 0.03\% & 6660770 & 0.03\% \\
\hline
  \end{tabular}
\end{table*}

\subsection{Calculation of 3C/3D ratio}

In this section, we are interested in computing the oscillator strength ratio of two of the brightest lines in Fe$^{16+}$, subject of a long-standing astrophysics puzzle that has recently been resolved~\cite{2022Fe16+}. In particular, we want to calculate the ratios of the oscillator strengths of the 3C [$2p^6$~$^1S_0 -2p^5 3d$~$^1P_1$] and 3D [$2p^6$~$^1S_0 - 2p^5 3d$~$^3D_1$] lines. 

Before calculating the matrix elements, we have to copy or symlink the even-parity \texttt{CONF.INP}, \texttt{CONF.DAT}, \texttt{CONF.DET}, and \texttt{CONF.XIJ} files as-is from the \texttt{/even} directory to the \texttt{/dtm} directory. From the \texttt{/odd} directory, we copy the \texttt{CONF.DET} and \texttt{CONF.XIJ} files to the \texttt{/dtm} directory, renaming them to \texttt{CONF1.DET} and \texttt{CONF1.XIJ}, respectively.  

The \texttt{pdtm} program takes in the input file \texttt{dtm.in}, where we can specify the parameters of the matrix element calculations. 
\begin{verbatim}
    # dtm.in
    Mode = TM
    Levels = 1 1, 4 5
    Operators = E1
\end{verbatim}
Here the first line \texttt{Mode = TM} indicates calculations of transition matrix elements. The second line tells the program to include the first energy level (\texttt{1 1}) from \texttt{CONF.XIJ}, and the fourth and fifth energy levels (\texttt{4 5}) from \texttt{CONF1.XIJ}. The third line creates an additional output file \texttt{E1.RES} with a summary of the resulting $E1$ matrix elements. 

The \texttt{pdtm} tabulates the $E1$ reduced matrix elements to the file \texttt{E1.RES}. Here, we obtain the reduced matrix elements $D(\mathrm{3D})=0.18416$ and $D(\mathrm{3C})=0.34540$. We can then compute the 3C/3D oscillator strength ratio
$$R(\mathrm{3C}/\mathrm{3D})=\left(\frac{D(\mathrm{3C})}{D(\mathrm{3D})}\right)^2\times \frac{\Delta E(\mathrm{3C})}{\Delta E(\mathrm{3D})}$$
where $\Delta E$ are the transition energies of the respective lines, to be 3.576. Despite using a small $12spdfg$ basis set, we find that our value agrees with recent experimental value of $3.51(2)_\mathrm{stat}(7)_\mathrm{sys}$~\cite{2022Fe16+}. A more accurate ratio can be obtained from advanced techniques described at the end of the previous section. 

\section{pCI-py scripts}\label{sec:pci-py}
In addition to the source codes distributed in the pCI package, we also include additional Python helper scripts to automate the entire computational workflow of pCI. These were written to improve the reproducibility of the results for the computations performed, as well as to make the overall pCI package more user-friendly. The included pCI-py scripts also serve as a template for more complex calculations, allowing users to customize them to fit their desired type of calculations. The included scripts require Python v3.x, and depend on a single user-defined YAML configuration file \texttt{config.yml}. 

Users can set the parameters defining their system of interest in \texttt{config.yml}, and run the various scripts to automate the respective pCI workflows. As an example, we will use a \texttt{config.yml} file used for automated computations of neutral Sr. The results of these calculations are used to provide data for the University of Delaware Portal for High-Precision Atomic Data and Computation. Refer to the pCI Read the Docs for the latest documentation and changes for these scripts.

The main scripts include:
\begin{enumerate}[noitemsep]
    \item \texttt{basis.py} - runs \texttt{hfd} and \texttt{bass}, and optionally, the CI+MBPT or CI+all-order programs.
    \item \texttt{ci.py} - runs \texttt{add}, \texttt{pbasc} and \texttt{pconf}.
    \item \texttt{dtm.py} - runs \texttt{pdtm}, and optionally, the programs to include the random phase approximation (RPA) correction to the matrix elements.
\end{enumerate}

There are also supplementary scripts that go beyond the scope of the pCI software package, including:
\begin{enumerate}[noitemsep]
    \item \texttt{upscale.py} - upscales the CI computation from a specified $nl$ basis to a larger $n^\prime l^\prime$ basis. Details of the logic of this script can be found in~\ref{sec:upscaling}.
    \item \texttt{isotope\_shifts.py} - automates isotope shift (IS) calculations from an already completed non-IS calculation.
    \item \texttt{gen\_portal\_csv.py} - compares results of CI computations to data from the NIST database, writing csv-formatted files. 
    \item \texttt{calc\_lifetimes.py} - generates csv-formatted data files of lifetimes and transition rates from output of \texttt{gen\_portal\_csv.py}
\end{enumerate}

\subsection{\texttt{config.yml}}
The \texttt{config.yml} file is a YAML-formatted configuration file that defines important parameters of the atomic system of interest. This configuration file is divided into several blocks:
\begin{itemize}[noitemsep]
    \item system parameters
    \item HPC parameters
    \item atom and method parameters
    \item parameters used by basis programs (only read by \texttt{basis.py})
    \item parameters used by ci programs (only read by \texttt{ci.py})
    \item parameters used by matrix element programs (only read by \texttt{dtm.py})
    \item parameters used by polarizability programs (only read by \texttt{pol.py})
    \item optional parameters defining types of calculations
\end{itemize}

The system parameters are in the \texttt{system} block and is read by all Python scripts:
\begin{verbatim}
    system:
        bin_directory: ""
        generate_directories: True
        run_codes: True
        on_hpc: True
        pci_version: default
\end{verbatim}
The fields
\begin{itemize}[noitemsep]
    \item \texttt{system.bin\_directory} specifies a path to a directory of pCI executable programs. If left blank, it is assumed the directories to the executable programs are in the user's environment PATH.
    \item \texttt{system.generate\_directories} specifies whether or not to generate directories for calculations. 
    \item \texttt{system.run\_codes} specifies whether or not the pCI programs should be run during the Python script execution. If set to \texttt{False}, the scripts will only generate the input files for the respective programs.
    \item \texttt{system.on\_hpc} specifies whether the user is running on an HPC environment. Setting this to \texttt{True} will create scripts that are submitted to the job scheduler (this is only compatible with the SLURM workload manager).
    \item \texttt{system.pci\_version} specifies which version of pCI to use from the HPC environment (this is relevant only for job scripts on HPC).
\end{itemize}

The atom parameters are in the \texttt{atom} block and is read by all scripts:
\begin{verbatim}
atom:
    name: Sr
    isotope:
    include_breit: True
    code_method: [ci+all-order, ci+second-order]
\end{verbatim}
The fields 
\begin{itemize}[noitemsep]
    \item \texttt{atom.name} specifies the name of the atomic system of interest. 
    \item \texttt{atom.isotope} specifies a specified isotope number. If left blank, the script will automatically find the atomic weight from the nuclear charge radii table of Ref.~\cite{nucrad}. This feature requires the \texttt{system.name} field to contain a valid atomic symbol.
    \item \texttt{atom.include\_breit} specifies in a boolean to set whether or not to include the Breit interaction in calculations.
    \item \texttt{atom.code\_method} specifies the calculation method to utilize. Available options are \texttt{ci}, \texttt{ci+all-order}, and \texttt{ci+second-order} for pure CI, CI+all-order, and CI+MBPT, respectively. Users can input \texttt{[ci+all-order, ci+second-order]} to run both CI+all-order and CI+MBPT calculations concurrently.
\end{itemize}

\subsection{\texttt{basis.py}}
The \texttt{basis.py} script automates the basis set construction for the CI computations, which is described in detail in Sec.~\ref{sec:basis}. By running the script, the basis set programs are run, resulting in \texttt{HFD.DAT}, which can be used by the next script for the CI calculations.
In addition to the \texttt{system} block, it also reads the \texttt{basis} block:
\begin{verbatim}
    basis:
        cavity_radius: 70
        diagonalized: False
        orbitals:
            core: 1s 2s 2p 3s 3p 3d 4s 4p 
            valence: 5s 5p 4d 6s 6p 5d 7s 7p 6d
            nmax: 35
            lmax: 5
        b_splines:
            nmax: 40
            lmax: 6
            k: 7
        val_aov:
            s: 5
            p: 5
            d: 5
            f: 3
        val_energies:
            kval: 1
            energies: 
                s: -0.28000
                p: [-0.22000, -0.22000]
                d: [-0.31000, -0.31000]
                f: [-0.13000, -0.13000]
\end{verbatim}
where the fields
\begin{itemize}
    \item \texttt{basis.cavity\_radius} specifies the size of the spherical cavity in a.u. 
    \item \texttt{basis.diagonalized} specifies whether to diagonalize the basis set or not.
    \item \texttt{basis.orbitals} specifies the core and valence orbitals to be included in the basis. It also requires a maximum principal quantum number \texttt{nmax} and maximum partial wave \texttt{lmax} for basis set orbital generation.
    \item \texttt{basis.b\_splines} specifies the maximum principal quantum number \texttt{nmax} (number of splines), maximum partial wave \texttt{lmax}, and order of the splines \texttt{k}.  
    \item \texttt{basis.val\_aov} specifies the number of valence orbitals to include for each partial wave in the all-order computations.  
    \item \texttt{basis.val\_energies} specifies the energies of the valence orbitals. \texttt{kval=1} is the default choice, where energies are set to the DHF energy of the lowest valence $n$ for the particular partial wave. In this case, the field \texttt{energies} can be safely ignored. \texttt{kval=2} allows specified energies of the valence orbitals, but is only used when the all-order valence energies are severely divergent. 
\end{itemize}

\texttt{basis.py} also reads the \texttt{optional} block:
\begin{verbatim}
    optional:
        qed:
            include: False
            rotate_basis: False
            
        isotope_shifts: 
            include: False
            K_is: 1
            C_is: 0.01
\end{verbatim}
where the fields
\begin{itemize}[noitemsep]
    \item \texttt{optional.qed} block allows users to specify the inclusion of QED corrections (this requires the QED package included in the \texttt{lib} directory). 
    \item \texttt{optional.isotope\_shifts} block allows users to specify isotope shift calculations by switching the \texttt{include} value to \texttt{True} and specifying keys \texttt{K\_is} and \texttt{C\_is}. These keys are described in more detail in~\ref{sec:isotope_shifts}.
\end{itemize}

The \texttt{basis.py} script also writes the standard output of the executables to their respective \texttt{*.out} files, e.g. \texttt{hfd} standard output is written to the file \texttt{hfd.out}. 

\subsection{\texttt{ci.py}}
The \texttt{ci.py} script automates the CI method, which is described in detail in Sec.~\ref{sec:ci}. It reads the \texttt{add.py} block for information about the configuration list:
\begin{verbatim}
    add:
        ref_configs:
            odd: [5s1 5p1]
            even: [5s2]
        basis_set: 17spdfg
        orbitals:
            core: 1s 2s 2p 3s 3p 3d 4s 4p 
            active: [
                4-7p:  0  4,
                4-7d:  0  4,
                4-7f:  0  4,
                5-7g:  0  4,
                ]    
        excitations:
            single: True
            double: True
            triple: False
\end{verbatim}
where the fields
\begin{itemize}
    \item \texttt{add.ref\_configs} requires a list of reference configurations to excite electrons from to construct the list of configurations defining the CI space. The list will not be constructed if left blank for a specified parity.
    \item \texttt{add.basis\_set} requires specification of the basis set designated by \texttt{nspdfg}, where \texttt{n} specifies the principal quantum number of the highest orbital allowed, and \texttt{spdfg} specifies the inclusion of the $s$, $p$, $d$, $f$, and $g$ orbitals. Higher partial waves can be included by appending to the end of the list \texttt{h}, \texttt{i}, \texttt{k}, etc.
    \item \texttt{add.orbitals} allows full customization of the allowed orbital occupancies. In the example, \texttt{4-7p: 0 4} defines the $(4-7)p$ shells to be open up to 4 occupancies. 
    \item \texttt{add.excitations} defines the types of excitations allowed by setting the sub-fields to be \texttt{True} or \texttt{False}.
\end{itemize}

Then it reads the \texttt{conf} block for parameters defining the CI execution itself:
\begin{verbatim}
    conf:
        odd:
            J: 1.0
            JM: 1.0
            J_selection: False
            num_energy_levels: 12
            num_dvdsn_iterations: 50
        even:
            J: 0.0
            JM: 0.0
            J_selection: False
            num_energy_levels: 24
            num_dvdsn_iterations: 50
        include_lsj: True
        write_hij: True
\end{verbatim}
where for each parity, the fields
\begin{itemize}[noitemsep]
    \item \texttt{conf.parity.J} defines the total angular momentum of the energy levels.
    \item \texttt{conf.parity.JM} defines the projection of the total angular momentum.
    \item \texttt{conf.parity.J\_selection} defines whether to calculate energy levels of a specified $J$ defined by \texttt{J} and \texttt{JM} or not.
    \item \texttt{conf.parity.num\_energy\_levels} defines the number of energy levels to be calculated for each parity.
    \item \texttt{conf.parity.num\_dvdsn\_iterations} defines the total number of Davidson iterations to allow.
\end{itemize}
and in general, the fields
\begin{itemize}[noitemsep]
    \item \texttt{conf.include\_lsj} defines whether the user wants the expectation values of the operators $L^2$ and $S^2$ to be calculated. 
    \item \texttt{conf.write\_hij} defines whether the user wants the Hamiltonian matrix to be written to the file \texttt{CONFp.HIJ}.
\end{itemize}
If \texttt{system.generate\_directories} or \texttt{system.run\_codes} is set to \texttt{True}, the \texttt{ci.py} script will generate directories for the CI calculations based on the parity and $J$ value specified. In the sample provided above, the directories \texttt{odd1} and \texttt{even0} will be generated for odd-parity calculations for $J=1$ levels and even-parity calculations with $J=0$ levels, respectively.

\subsection{\texttt{dtm.py}}
The \texttt{dtm.py} script automates the matrix element calculations, which are described in detail in Sec.~\ref{sec:dtm}. Specifically, this script prepares directories for calculations with \texttt{pdtm} by moving the relevant input files from previous calculations to them. This script reads the \texttt{dtm} block:
\begin{verbatim}
    dtm:
        include_rpa: True
        DM: 
            matrix_elements: 
            level_range: 
                odd: 
                even: 
        TM:
            matrix_elements: E1
            from:
                parity: odd
                level_range: 1 3
            to:
                parity: even 
                level_range: 1 1
\end{verbatim}
where the fields
\begin{itemize}[noitemsep]
    \item \texttt{dtm.include\_rpa} defines whether the user would like to include RPA corrections. Note that this option requires compilation of the \texttt{rpa} and \texttt{rpa-dtm} programs. 
    \item \texttt{dtm.DM} defines job parameters for density matrix (DM) calculations, while \texttt{dtm.TM} defines job parameters for transition matrix (TM) calculations.
    \item \texttt{dtm.*.matrix\_elements} defines operators to include RPA if \texttt{include\_rpa = True}. This value can be a single matrix element or an array of matrix elements. In addition, a separate file \texttt{*.RES} will be generated by \texttt{pdtm} for each operator listed here. For \texttt{DM}, the operators include: \texttt{GF}, \texttt{A\_hf} and \texttt{B\_hf}. For \texttt{TM}, the operators include: \texttt{E1\_L}, \texttt{E1\_V}, \texttt{E1}, \texttt{E2}, \texttt{E3}, \texttt{M1}, \texttt{M2}, \texttt{M3}, \texttt{EDM}, \texttt{PNC}, \texttt{AM}, \texttt{MQM}. To include multiple operators, list the operators in brackets, e.g. ``\texttt{[E1, M2]}''.
    \item \texttt{dtm.*.level\_range} defines the level ranges to calculate matrix elements. The indices of each energy level can be found in the file \texttt{FINAL.RES} or \texttt{CONF.RES} generated by the \texttt{pconf} program. DM calculations require one of the two parities (\texttt{odd} or \texttt{even}) to be filled with a level range. TM calculations require all fields \texttt{TM.from.*} and \texttt{TM.to.*} to be filled with the respective parities, as well as the initial and final level ranges of the transitions.    
\end{itemize}

\subsection{\texttt{pol.py}}
The \texttt{pol.py} script automates polarizability calculations, which are described in detail in Sec.~\ref{sec:pol}. This script reads the \texttt{pol} block:
\begin{verbatim}
    pol:
        parity: even
        level: 1
        method: 1
        field_type: static, dynamic
        wavelength_range: 1000 1000
        step_size: 0
\end{verbatim}
where the fields
\begin{itemize}[noitemsep]
    \item \texttt{pol.parity} specifies the parity of the state to calculate polarizabilities for.
    \item \texttt{pol.level} specifies the index of the state to calculate polarizabilities for.
    \item \texttt{pol.method} specifies the method to calculate polarizabilities. This field corresponds to the integer parameter \texttt{Method} described in Sec.~\ref{sec:pol}.
    \item \texttt{pol.field\_type} specifies calculations of static or dynamic polarizabilities. Calculations of both can be done by setting this field to \texttt{[static, dynamic]}.
    \item \texttt{pol.wavelength\_range} specifies the range of wavelengths to calculate dynamic polarizabilities for.
    \item \texttt{pol.step\_size} specifies the step size of the range of wavelengths.
\end{itemize}

\subsection{Running the pCI-py scripts}
After configuring the \texttt{config.yml} file, running the CI computations becomes as simple as running 2 Python scripts: \texttt{basis.py} and \texttt{ci.py}. If matrix elements are required, then the additional \texttt{dtm.py} script can be run. If polarizabilities are required, then the \texttt{pol.py} script can be run as well.

Two sample \texttt{config.yml} files are included in the pCI software package. The \texttt{config\_Fe16+.yml} automates the pure-CI method for the highly charged ion Fe$^{16+}$, while the \texttt{config\_Sr.yml} automates the CI+all-order and CI+second-order calculations to produce a large volume of atomic data used by the University of Delaware Portal for High-Precision Atomic Data and Computation. 

\section{Atomic properties of neutral Sr}\label{sec:sr}
In this section, we demonstrate the usage of the pCI-py scripts described in the previous section to calculate energies, reduced $E1$ matrix elements, and static and dynamic polarizabilities of neutral Sr for the $^1S_0$ state. Here, we will use the file \texttt{config\_Sr.yml} sample configuration file included in the pCI distribution, which is displayed in parts in Sec.~\ref{sec:pci-py}. 

We treat Sr as a divalent ion, constructing the basis set in the $V^{N-2}$ approximation, where $N$ is the number of electrons. The \texttt{basis.py} script will construct a basis set in both CI+all-order and CI+MBPT approximations, in their respective directories. Both are constructed as non-diagonalized basis sets in a spherical cavity radius of 70 a.u. The initial HFD self-consistency procedure is carried out for the core $1s$, $2s$, $2p$, $3s$, $3p$, $3d$, $4s$, and $4p$ orbitals, and then the valence $5s$, $5p$, $4d$, $6s$, $6p$, $5d$, $7s$, $7p$, and $6d$ orbitals are constructed in the frozen-core potential. The remaining virtual orbitals are formed using 40 B-spline orbitals of order 7, including partial waves up to $l=5$. The coupled-cluster equations are then solved to all-order and to second-order in the respective directories. These basis set is stored in the file \texttt{HFD.DAT}, while corrections from all-order and MBPT are stored in the form of effective radial integrals in the files \texttt{SGC.CON} and \texttt{SCRC.CON}, respectively. 

Next, the \texttt{ci.py} script generates lists of configurations by exciting electrons from the odd-parity $5s 5p$ and even-parity $5s^2$ configurations to all orbitals up to $17spdfg$, then run the CI computations. The energies and wave functions of the lowest 12 states with $J=1$ are calculated for the odd-parity CI run, and the lowest 24 with $J=0$ for the even-parity CI run. 

The \texttt{dtm.py} script is then run to calculate $E1$ reduced matrix elements, including RPA corrections. In \texttt{config\_Sr.yml} we specify transitions from the first 3 odd states to the first even state. We present the results of these computations for $^3P_1^o-\,^1S_0$ and $^1P_1^o-\,^1S_0$ in Table~\ref{tab:sr_e1}, and compare them with previous theory~\cite{Safronova2013} and experimental~\cite{Sr1997} values. 

\begin{table}
\centering
\caption{\label{tab:sr_e1} Comparison of the reduced electric-dipole matrix elements (in a.u.) obtained from pCI with previous theory~\cite{Safronova2013} and experimental~\cite{Sr1997} values. The values displayed under ``pCI'' are obtained from the CI+all-order approach, with uncertainty calculated as the difference between CI+all-order and CI+MBPT values.}
\begin{tabular}{lcccc}
\hline
\multicolumn{1}{c}{Transition}&
\multicolumn{1}{c}{pCI}&
\multicolumn{1}{c}{Theory~\cite{Safronova2013}}&
\multicolumn{1}{c}{Expt.~\cite{Sr1997}} \\
\hline \\ [-0.8pc]
$^3P_1^o$ -- $^1S_0$ & 0.155(4) & 0.158 & 0.151(2) \\
$^1P_1^o$ -- $^1S_0$ & 5.275(19) & 5.272 & 5.248(2) \\
\hline
  \end{tabular}
\end{table}

Finally, the \texttt{pol.py} script is run to calculate dc and ac valence polarizabilities for the $^1S_0$ state. We present these results in Table~\ref{tab:sr_pol}, along with that of $^3P_0^o$ and the difference $\alpha(^3P_0^o)-\alpha(^1S_0)$. These results are compared with previous theory results from Ref.~\cite{Safronova2013}, which uses the CI+all-order+RPA approach with a different basis set. Note that only the valence polarizabilities are compared, with core-core and core-valence contributions neglected. The final column displays ac polarizabilities for each state at 1000 nm. To calculate polarizabilities for the $^3P_0^o$ state, one has to set \texttt{pol.parity} to \texttt{odd}, then swap the values of \texttt{conf.odd.J} and \texttt{conf.odd.JM} with \texttt{conf.even.J} and \texttt{conf.even.JM}. The \texttt{ci.py} script is re-run to calculate wave functions in the respective projections, and then \texttt{pol.py} is re-run to obtain polarizabilities. 

As an extra test, one can calculate valence polarizabilities for the $^3P_1^o$ state by setting \texttt{conf.odd.J} and \texttt{conf.odd.JM} to \texttt{1}, \texttt{conf.even.J} and \texttt{conf.even.JM} to \texttt{0}, and re-running \texttt{pol.py}. Doing so, we obtain scalar polarizability $\alpha_0(^3P_1)=194.84$ a.u., vector polarizability $\alpha_2(^3P_1)=23.45$ a.u., and total ac polarizability $\alpha(^3P_1)=218.29$ a.u. at $\lambda=1000$ nm. 
Higher precision of the polarizabilities can be attained by including core polarizabilities, which are calculated with a different program not included in this work.

\begin{table}
\centering
\caption{\label{tab:sr_pol} Valence polarizabilities of Sr states $\alpha_0(5s^2\,^1S_0)$, $\alpha_0(5s5p\,^3P_0^o)$, and their difference, calculated with the pCI programs using the CI+all-order approach. The columns ``pCI'' and ``Theory'' display dc polarizabilities, while the final column ``ac'' displays calculated ac polarizabilities at 1000 nm. dc results are in a.u., and compared with previous theory values of Ref.~\cite{Safronova2013}. }
\begin{tabular}{cccc}
\hline
\multicolumn{1}{c}{State}&
\multicolumn{1}{c}{pCI}&
\multicolumn{1}{c}{Theory~\cite{Safronova2013}}&
\multicolumn{1}{c}{ac}\\
\hline \\ [-0.8pc]
$5s^2\,^1S_0$   & 192.78 & 192.51 & 242.97 \\
$5s5p\,^3P_0^o$ & 455.06 & 452.55 & 186.34 \\
$^3P_0^o-{}^1S_0$ & 262.28 & 260.30  \\
\hline
  \end{tabular}
\end{table}

\section{Acknowledgements}
\label{sec:acknowledgements}
We thank Dr. Jeffrey Frey from the University of Delaware IT-RCI group for helpful discussions and contributions to optimizing parts of the \texttt{pconf} program. 
The developments and calculations in this work were done through the use of IT resources at the University of Delaware, specifically the high-performance Caviness and DARWIN computer clusters. 
This work was supported by the US NSF Grants No. OAC-2209639 and PHY-2309254. 
MK thanks Russian Science Foundation, grant \# 23-22-00079. AIB contributed to this work prior to February 2022.


\bibliographystyle{elsarticle-num}
\bibliography{refs}

\newpage

\appendix

\section{CI+X Extensions}
The capabilities of the pCI software package can be extended to use the CI+all-order and CI+MBPT methods. The all-order and MBPT methods include corrections to the bare Hamiltonian due to core shells, which can then be accounted for in the CI. The pCI software package includes an older Fortran 77 version of the all-order package~\cite{CIall}, while a more modern Fortran 90 version is set to be released at a later time. The Read the Docs page will be updated following the release of the modern all-order package. A pCI-compatible MBPT package can be installed following Ref.~\cite{KozPorSaf15}. 

\subsection{CI+all-order}
The all-order label refers to the inclusion of a large number of terms (second-, third-, fourth-order, etc.) in order-by-order many-body perturbation theory expansion using the iterative solutions until sufficient numerical convergence is achieved. The included all-order package implements a variant of the linearized coupled-cluster single double (LCCSD) method. This version of CC has been developed specifically for atoms fully utilizing atomic symmetries and is capable of being efficiently run with very large basis sets (over 1000 orbitals), reaching negligible numerical uncertainty associated with the choice of basis set. Formulas can be found in Ref.~\cite{CIall}.

The all-order package consists of four codes: \texttt{allcore-ci}, \texttt{valsd-ci}, \texttt{sdvw-ci}, and \texttt{second-ci}. The \texttt{allcore-ci}, \texttt{valssd-ci}, and \texttt{sdvw-ci} programs calculate core-core, core-valence, and valence-valence excitations, respectively. The \texttt{second-ci} program calculates corrections in second-order MBPT, but for a much larger part of the Hamiltonian than the all-order codes, since high accuracy is not required for corrections associated with higher orbitals. 

If the all-order calculation was carried out, it will overwrite the second-order results with the all-order results where available. Such overlay of the MBPT and the all-order parts drastically improves the efficiency of the method. Note that \texttt{second-ci} can also be run as a standalone program, replacing the MBPT programs of the 2015 CI-MBPT. These codes store the resulting data in \texttt{SGC.CON} and \texttt{SCRC.CON}. The \texttt{SGC.CON} file is typically small, whereas \texttt{SCRC.CON} can take up to a few GB in size. 

Table~\ref{tab:ao_io_files} displays a list of the all-order programs and their input and output files. The runtimes of the all-order programs are dependent on the atom size, from a few minutes to nearly a day for the included F77 version.

\begin{table*}[ht]
    \centering
    \begin{threeparttable}
    \caption{\label{tab:ao_io_files} List of the all-order programs and their input and output files. The columns ``Program'' and ``Description'' lists the names of the programs and their main capability, respectively. The column ``Input (text)'' lists user-defined text input files. The column ``Input (binary)'' lists binary input files that are themselves constructed from a program (the \texttt{SGC.CON} file is a text file). All programs write their results to standard output. The pCI-py scripts write outputs of the all-order programs to \texttt{out.*}. }
    \begin{tabular}{ccccc}
        \hline \\[-3mm]
        Program & Description & Input (text) & Input (binary) & Output \\[1mm]
        \hline \\[-3mm]
         \texttt{tdhf/bdhf$^a$}  & Solves HFD equations & \texttt{bas\_wj.in}$^b$ & & \texttt{fort.1}  \\[2mm]
         \texttt{nspl/bspl$^a$} & Produces B-spline basis & & \texttt{fort.1} & \texttt{hfspl.1}  \\
           &   & &   & \texttt{hfspl.2}  \\[2mm]
         \texttt{bas\_wj}  & Converts B-spline basis to pCI basis format & & \texttt{hfspl.1} & \texttt{WJ.DAT}  \\
           &   & &  \texttt{hfspl.2} &   \\[2mm]
         \texttt{bas\_x} & Converts pCI basis to all-order format &  & \texttt{HFD.DAT} & \texttt{hfspl.1}   \\
           &   & &  & \texttt{hfspl.2}  \\[2mm]
         \texttt{allcore-ci} & Calculates core-core excitations & \texttt{inf.aov}$^b$ &  \texttt{hfspl.1} & \texttt{pair.3}  \\ 
         & & & \texttt{hfspl.2} &     \\ [2mm]
         \texttt{valsd-ci} & Calculates core-valence excitations & \texttt{inf.aov}$^b$ & \texttt{hfspl.1} &  \texttt{val2}  \\
         & & & \texttt{hfspl.2} & \texttt{sigma}     \\ 
         & & & \texttt{pair.3} &    \\ [2mm]
         \texttt{sdvw-ci} & Calculates valence-valence excitations & \texttt{inf.aov}$^b$ & \texttt{hfspl.1} & \texttt{pair.vw}  \\
         & & & \texttt{hfspl.2} &  \texttt{sigma1}   \\
         & & & \texttt{val2} &   \\ 
         & & & \texttt{pair.3} &   \\ [2mm]
         \texttt{second-ci} & Calculates second-order MBPT corrections & \texttt{inf.vw}$^b$ & \texttt{hfspl.1} & \texttt{SGC.CON} \\
         & & & \texttt{hfspl.2} & \texttt{SCRC.CON}   \\
         & & & \texttt{sigma1}  \\
         & & & \texttt{pair.vw}  \\ 
         & & & \texttt{HFD.DAT}  \\ [2mm]
         \hline
    \end{tabular}
    \begin{tablenotes}
       \item $^a$ Program starting with ``b'' includes Breit corrections.
       \item $^b$ Program reads these files as standard input. These are the input files generated by the \texttt{basis.py} script.
    \end{tablenotes}
    \end{threeparttable}
\end{table*}

\subsection{Basis sets for CI+all-order}
For the CI+all-order and CI+MBPT methods, basis sets are constructed using a combination of HFD orbitals and B-splines to accommodate a large number of orbitals. However, one needs too many B-splines to reproduce core and lower valence stats with high enough accuracy for heavier atoms. Therefore, the core and a few valence electron wave functions are taken from Dirac-Hartree-Fock (DHF), and a combined basis with splines is built. More splines will mean a larger basis for the CI as well, so we would like to avoid this. As the CI and all-order code packages were originally developed separately, they use basis sets in different formats. The CI programs use \texttt{HFD.DAT} and all-order programs use \texttt{hfspl.1} and \texttt{hfspl.2} files. The program \texttt{bas\_wj} converts the all-order formatted \texttt{hfspl.1} and \texttt{hfspl.2} to the CI formatted \texttt{HFD.DAT} format, while the program \texttt{bas\_x} converts them the other way around. Note that the \texttt{bas\_wj} program produces the \texttt{HFD.DAT} formatted file called \texttt{WJ.DAT}. 

The construction of the basis set for CI+all-order or CI+MBPT calculations follows the general recipe:
\begin{enumerate}[noitemsep]
    \item Produce B-splines
    \begin{enumerate}
        \item \texttt{tdhf < bas\_wj.in} - solves HFD equations in all-order format
        \item \texttt{bspl40 < spl.in} - produces B-spline basis
    \end{enumerate}
    \item \texttt{hfd} - solves DHF equations in CI format
    \item \texttt{bas\_wj} - converts B-spline basis to CI basis format (\texttt{WJ.DAT})
    \item \texttt{bass} - build combined basis from \texttt{HFD.DAT} and \texttt{WJ.DAT}
    \item \texttt{bas\_x} - convert combined basis from CI format to all-order format
    \item Run all-order or MBPT codes
\end{enumerate}

The Python scripts described in Sec.~\ref{sec:pci-py} follow the above recipe to construct the basis set and run the all-order and MBPT programs. A more detailed description of this recipe and the corresponding programs can be found on the pCI Read the Docs.

\subsection{CI+MBPT}
See Ref.~\cite{KozPorSaf15} for details on the CI-MBPT package. 

\section{Upscaling the basis set}\label{sec:upscaling}
The typical endpoint of the pCI workflow is to converge the energies obtained from the \texttt{pconf} program for the basis set. This may involve many computations with increasing principal quantum number $n$ or higher partial waves $l$. However, users may find that their next CI computation will become intractable on their computing system. In this case, users can employ a technique we call \textit{upscaling the basis set}. This involves reducing the size of the CI space of a previous calculation by selecting only the most important configurations and removing all others from the list. We define the importance of a configuration $\Psi_i$ by their weight $w_i=\sum_{j=1}^{N_i}|c_{i,j}|^2$, where $N_i$ is the number of determinants in configuration $i$. A logarithmic cutoff $x$ is chosen by balancing the number of configurations obtained by removing those below the cutoff with the subsequent energy difference between the direct run and the cut run. $x$ is typically chosen so that the resulting energy difference is minimal for any resulting energy level. This energy difference is then subtracted from the results of subsequent CI calculations involving higher $n$ or $l$. 

This requires the \texttt{con\_cut} and \texttt{merge\_ci} codes of the CI-MBPT code package. See Ref.~\cite{KozPorSaf15} for information on how to install them.

\section{RPA corrections}
The optional \texttt{RPA} programs calculate RPA corrections to the matrix elements. First, the \texttt{rpa} program solves RPA equations, calculates radial integrals of the effective operators and writes them to files \texttt{RPA\_n.INT}, where the \texttt{n=1-13} numerates one-electron operators in the same order as they are listed in the block \texttt{RPA} of the file \texttt{MBPT.INP}. This program reads three blocks of the file \texttt{MBPT.INP}, namely \texttt{MBPT}, \texttt{Eval}, and \texttt{RPA}. 

After \texttt{rpa} is complete, the \texttt{rpa\_dtm} program rewrites the radial integrals from the \texttt{RPA\_n.INT} files to the \texttt{DTM.INT} file, which can subsequently be used by the \texttt{pdtm} program to calculate matrix elements. In this way, we take into account the RPA corrections and find matrix elements of the effective operators. Note that before running \texttt{rpa\_dtm}, the file \texttt{DTM.INT} has to be constructed by the \texttt{pdtm} program. This is simply done by running \texttt{pdtm} using the \texttt{Mode=Init} option. 
More details of the programs \texttt{rpa} and \texttt{rpa\_dtm} can be found in Ref.~\cite{KozPorSaf15}. 

Once the RPA corrections have been inserted into \texttt{DTM.INT}, \texttt{pdtm} has to be run again. This time, the program calculates matrix elements between specified energy levels with RPA corrections. 

\section{QED corrections}\label{sec:qed}
The optional QED programs include QED corrections in the CI calculations in the form of effective radial integrals via the file \texttt{SGC.CON}, similar to how MBPT and all-order corrections are included. Inclusion of QED corrections require three extra programs, \texttt{sgc0}, \texttt{qed\_rot} and \texttt{qedpot\_conf}, which can be compiled in the \texttt{lib} directory. 

The \texttt{sgc0} program is first run to generate an empty \texttt{SGC.CON} file. 
The \texttt{qedpot\_conf} program uses the empty \texttt{SGC.CON} file as a template, and writes QED corrections to the file \texttt{SGCqed.CON} as one-electron effective radial integrals, if they are nonzero. 
%The program also produces the file \texttt{uehling.dat}, which contains Uehling vacuum-polarization potential, and the file \texttt{wk.dat}, which contains Wichmann-Kroll point nuclear vacuum-polarization potential. 

The \texttt{qed\_rot} program rotates the orbitals by diagonalizing the HFD Hamiltonian using the one-electron QED matrix elements calculated by \texttt{qedpot\_conf}. This process iterates until the change in the core orbitals converges.

By default, \texttt{qedpot\_conf} reads the basis functions from the file \texttt{CONF.DAT} and uses the QEDMOD potential~\cite{Shabaev2013} for calculating the matrix elements. Other options can be specified in the input file \texttt{qedpot.inp}, where the file for basis functions (e.g., \texttt{HFD.DAT}) and the key \texttt{kvar}, which determines the variant of the QED potential used, should be provided.
%The input parameters for \texttt{qedpot\_conf} are the file, where the basis functions are read from: usually \texttt{HFD.DAT} or \texttt{CONF.DAT} (default) and the key \texttt{kvar}, which determines the variant of the QED potential used. 
The possible values of this key are described here:
\begin{itemize}[noitemsep]
    \item \texttt{1} - QEDMOD from Ref.~\cite{Shabaev2013} (default value)
    \item \texttt{2} - Self-energy (SE) local potential from Ref.~\cite{Flambaum2005} + non-local correction from Ref.~\cite{Shabaev2013} 
    \item \texttt{3} - SE local potential from Ref.~\cite{Flambaum2005}
    \item \texttt{4} - QEDPOT from Ref.~\cite{Tupitsyn2013}
    \item \texttt{5} - Semi-empirical potential from Ref.~\cite{Johnson1985}
\end{itemize}

The input parameters for \texttt{qed\_rot} are
\begin{itemize}[noitemsep]
    \item \texttt{Kdg} - (1 - general diagonalization, 2 - first-order diagonalization)
    \item \texttt{Ksg} - (1 - do not include QED, 2 - include QED)
    \item \texttt{Kbrt} - (0 - no Breit, 1 - Gaunt, 2 - Full Breit)
\end{itemize}

The \texttt{optional.qed} block of the \texttt{basis.py} script allows users to specify inclusion of QED corrections when constructing the basis set. It generates a bash script \texttt{batch.qed}, which runs the programs \texttt{qedpot\_conf} and \texttt{qed\_rot} and creates the corresponding input files. 

\section{Isotope shifts}\label{sec:isotope_shifts}

To run isotope shift calculations, the entire pCI workflow is essentially repeated several times, varying the key \texttt{C\_is}, which defines the isotope shift perturbation. The type of isotope shift calculation is defined by the key \texttt{K\_is}. For specific mass shift calculations, there is an additional key \texttt{Klow}, which determines the method for calculating the matrix elements of the Hamiltonian. These keys are set in the input files \texttt{HFD.INP}, \texttt{BASS.INP} and \texttt{CONF.INP}. 
These keys are described here:
\begin{itemize}[noitemsep]
    \item \texttt{K\_is} defines the type of isotope shift calculation
    \begin{itemize}
        \item \texttt{0} - no isotope shift
        \item \texttt{1} - field shift
        \item \texttt{2} - specific mass shift
        \item \texttt{3} - normal mass shift
        \item \texttt{4} - total mass shift (sum of \texttt{K\_is=2} and \texttt{K\_is=3})
    \end{itemize}
    \item \texttt{C\_is} defines the isotope shift perturbation 
    %$c_\mathrm{IS}=dR_\mathrm{nucl}/R_\mathrm{nucl}$ \ab{This is valid only for the field shift.}
    \item \texttt{Klow} defines the method for calculating the matrix elements of the mass-shift Hamiltonian
    \begin{itemize}
        \item \texttt{0} - nonrelativistic form of the Hamiltonian, the lower component of the Dirac bispinor is not taken into account
        \item \texttt{1} - nonrelativistic form of the Hamiltonian, the lower component of the Dirac bispinor is taken into account
        \item \texttt{2} - relativistic form of the Hamiltonian, the lower component of the Dirac bispinor is taken into account
    \end{itemize}
\end{itemize}

Note that for sufficient accuracy of the isotope shift calculations, it is important to compile the programs \texttt{pbasc} and \texttt{pconf} with arrays of two-electron radial integrals set to the \texttt{real(double precision)} datatype, rather than the default \texttt{real(single precision)}. This can be done by changing the value of the \texttt{type2\_real} parameter in the \texttt{params.f90} file from \texttt{sp} to \texttt{dp}. A correct compilation of this can be confirmed by looking at the title of the output files \texttt{BASC.RES} and \texttt{CONF.RES}, which should mention ``\texttt{double precision for 2e integrals}''.

Repeating the entire pCI workflow manually can be very time-consuming and error-prone. Instead, users can use the pCI-py scripts to automate isotope shift calculations. This is done simply by setting the \texttt{optional.isotope\_shifts} block in the \texttt{config.yml} file.
\begin{verbatim}
    optional:
        isotope_shifts:
            include: True
            K_is: 1
            C_is: 0.01
\end{verbatim}
By setting \texttt{optional.isotope\_shifts.include} to \texttt{True}, the script will automatically generate directories for isotope shift calculations in the stage \texttt{basis.py}. For the sample above, the \texttt{basis.py} script will generate the following file directory:
\begin{verbatim}
    config.yml
    basis.py
    ...
    /IS
        /minus0.01
        /minus0.005
        /0
        /plus0.005
        /plus0.01
\end{verbatim}

Repeating with \texttt{K\_is=2} will generate the \texttt{SMS} directory in the root directory, etc. 

Typically, one starts with a normal CI computation without isotope shifts to check the accuracy of energy levels or other properties before starting isotope shift calculations. In this case, users can use the \texttt{isotope\_shifts.py} script to change the \texttt{K\_is}, \texttt{C\_is} and \texttt{Klow} values from some previously completed \texttt{0} directory. This script simply takes the input files and changes the values of the relevant keys and reruns the programs. The \texttt{isotope\_shifts.py} program asks the user to input the \texttt{K\_is} and \texttt{C\_is} values at runtime, then asks if they would like to generate the basis set (\texttt{basis}), run the CI calculation (\texttt{ci}) or analyze the resulting data (\texttt{analysis}). The \texttt{analysis} option retrieves the energy levels from each isotope shift directory, then creates a csv file with each energy level and its calculated isotope shift value. The script calculates field shifts in units GHz/fm$^2$, and mass shifts in units GHz$\cdot$amu. 






% \end{thebibliography}

\end{document}